\documentclass[12pt]{iopart}   

\usepackage{iopams}    
\usepackage{amsfonts}   
\usepackage{amssymb}
\usepackage{graphicx}   
\usepackage{subfigure}
\newcommand{\argmin}{\rm arg\,min}
\begin{document}

\maketitle

\title[Multilevel Discretized Random Fields and Spatial Simulations]
{Multilevel Discretized Random Field Models with ``Spin''
Correlations for the
 Simulation of Environmental Spatial Data}
\author{Milan \v{Z}ukovi\v{c}}
 \address{Technical University of Crete, Geostatistics Research Unit,
Chania 73100, Greece}
 \ead{mzukovic@mred.tuc.gr}   
\author{Dionissios T. Hristopulos}
 \address{Technical University of Crete, Geostatistics Research Unit,
Chania 73100, Greece}
 \ead{dionisi@mred.tuc.gr}   
\date{\today}
\begin{abstract}
A current problem of practical significance is the analysis of
large, spatially distributed,  environmental data sets. The problem
is more challenging for variables that follow non-Gaussian
distributions.  We show by means of numerical simulations that the
spatial correlations between variables can be captured by
interactions between ``spins''. The spins represent multilevel
discretizations of environmental variables with respect to a number
of pre-defined thresholds. The spatial dependence between the
``spins'' is imposed by means of short-range interactions. We
present two approaches, inspired by the Ising and Potts models, that
generate conditional simulations of spatially distributed variables
from samples with missing data. Currently, the sampling and
simulation points are assumed to be at the nodes of a regular grid.
The conditional simulations of the ``spin system'' are forced to
respect locally the sample values and the system statistics
globally. The second constraint  is enforced by minimizing a cost
function representing the deviation between normalized correlation
energies of the simulated and the sample distributions. In the
approach based on the $N_c-$state Potts model, each point is
assigned to one of $N_c$  classes. The interactions involve all the
points simultaneously. In the Ising model approach, a sequential
simulation scheme is used: the discretization at each simulation
level is binomial (i.e., $\pm 1$.) Information propagates from lower
to higher levels as the simulation proceeds. We compare the two
approaches in terms of their ability to reproduce the target
statistics (e.g., the histogram and the variogram of the sample
distribution), to predict data at unsampled locations, as well as in
terms of their computational complexity. The comparison is based on
a non-Gaussian data set (derived from a digital elevation model of
the Walker lake area, Nevada, USA). We discuss the impact of
relevant simulation parameters, such as the domain size, the number
of discretization levels, and the initial conditions.
\end{abstract}

  \submitto{Journal of
Statistical Mechanics: theory and experiment}


\maketitle

\section{Introduction}
\label{intro}

Spatially distributed data are common in the physical sciences. They
represent various environmental variables, such as contaminant
concentrations in the atmosphere,  soil permeability and
dispersivity, etc. To account for the spatial variation and the
measurement uncertainty of such quantities, the mathematical model
of spatial random fields is commonly used. To date, huge numbers of
large spatial data sets are gathered from around the globe on a
daily basis. The efficient storage, analysis, harmonization, and
integration of such data is of great importance in various
scientific areas such as image processing, pattern recognition,
remote sensing, and environmental monitoring. Often the data
coverage is incomplete for various reasons, such as meteorological
conditions (measurements hindered by clouds, aerosols, or heavy
precipitation) or equipment limitations (values below detection
level or resolution threshold). Hence, in order to use standard
tools for the data analysis and visualization, one needs to deal
with the problem of incomplete (missing) data. In addition, data
sampled at different resolutions may need to be combined (e.g., in
data fusion methods). This requires down-scaling (refining) of the
data with the coarser resolution.

These tasks can be performed by means of well established
interpolation and classification techniques~\cite{atk99}. While
deterministic methods (e.g., nearest-neighbor or inverse distance
interpolation) can be used, stochastic methods are often preferred
because they are more flexible in incorporating spatial correlations
and and they provide estimates of prediction uncertainty. However,
considering the ever-increasing size of spatial data, stochastic
methods such as kriging~\cite{wack03}, can be impractical due to
their high computational complexity, requiring the use of high
performance computational technologies~\cite{hawi98}. Furthermore,
kriging is founded on the assumption of a jointly Gaussian
distribution, which is often failed by the data. Practical
application of kriging involves considerable human (subjective)
input, regarding the selection of the correlation (variogram) model
and the kriging neighborhood selection~\cite{dig07}.

The classical geostatistical approach relies on the structure
function (variogram) for modeling the spatial correlations. However,
the correlations can also be considered as the outcome of
``interactions'' between the field values at different
points~\cite{dth03} that generate short-range spatial order. In the
case of static disorder, which is typical in case of ``quenched''
geological disorder or measurements representing a single time slice
of a dynamical process, such interactions represent ``effective
constraints'' imposed by the underlying process. For example, in the
case of a digital elevation model, these constraints are imposed by
the topography of the area. While the nature of the constraints may
differ significantly between processes, we believe that  universal
aspects of spatial correlations can be captured by means of
relatively simple effective interactions. By incorporating the
interactions in an energy functional a Gibbs probability measure can
be defined as in~\cite{dth03}. The realization probability of each
spatial configuration is then governed by the relative weight of the
``many body'' Gibbs probability density function. In the Gaussian
case it is straightforward to define the interactions between the
random field values at neighboring locations. On a regular lattice
this approach defines Gauss-Markov random fields.

In the non-Gaussian case, it is convenient to discretize the
continuous values of the random field using a set of discrete
``spins''. We arbitrarily define higher spin values to represent
higher values of the field. In the spirit described above, one can
then consider interactions between different ``spins''. In this
framework, the problems of spatial interpolation and simulation are
mapped into a spatial classification problem, in which each
``prediction'' location is assigned to a specific spin value
(class). The concept of classical spins is a suitable choice for
modeling the multilevel discretization of the continuous field. Spin
models from statistical physics, e.g., the Ising and Potts models,
have already been applied in various problems in economy and finance
\cite{man_stan99,bo_pot00,bouch00}, materials science
\cite{kard95,wouts07,zhen06}, and biology \cite{gran92}. However,
these studies  focus mainly on long-range correlations. In the
 framework of Gibbs-Markov random fields, the short-range correlation
properties of Potts and Ising models have been widely applied in
image analysis (see e.g.
\cite{besa86,gimel99,kato06,zhao07,tana95}). The Potts model in
super-paramagnetic regime was also proposed as a data clustering
model \cite{blatt96}. We introduce here two non-parametric models
for spatial classification that are loosely based on the Ising and
Potts spin models. The first model employs a sequential
classification approach while the second a simultaneous (parallel)
classification of all levels.


The rest of the paper is organized as follows.  In Section
\ref{spat_pred}, we present the problem of spatial
classification/prediction, and we review the commonly used
classification algorithm of $k$-nearest neighbors. Section
\ref{models_nnc} briefly reviews the Ising and Potts models and
introduces the ``spin'' nearest-neighbor correlation models that we
propose for spatial classification. In Section \ref{simul} we
investigate technical aspects of the simulations, and we describe
computational details. Section \ref{results} focuses on the case
study: it presents computational details as well as the analysis of
the simulation results. Finally, in Section \ref{conclusions}, we
summarize the relevant results and point out some future directions.

\section{Spatial prediction and classification}
\label{spat_pred}

Let us consider a set of sampling points $G_{s}=\{\vec{r}_{i}\} $,
where $\vec{r}_{i}=(x_i, y_i) \in {\mathbb{R}}^2$ and
$i=1,\ldots,N$. These points are assumed to be scattered on a
rectangular grid $\tilde{G}$ of size $N_{\tilde{G}}=L_{x} \, L_{y} >
N$, where $L_{x}$ and $L_{y}$ are respectively the number of nodes
in the orthogonal directions (in terms of the unit length). If
$z_{i}$ is a value attributed to the point $\vec{r}_{i}$, the set
$Z\{G_{s}\}=\{z_{i} \in {\mathbb{R}} \}$ represents the sample of
the process.  Here we assume that $Z\{G_{s}\}$ represents a sample
from a realization of a continuous random field ${\rm
Z}({\vec{r}};\omega),$ where $\omega$ is the state index.

Let $G_{p}=\{\vec{r}_{p}\} $ be the set of prediction points where
$p=1,\ldots, P$, such that $\tilde{G}=G_{s} \cup G_{p}$.   We
discretize the continuous distribution of $Z$ using a number of
classes, ${\mathcal C}_q, \, q=1,\ldots,N_c$.  The classes are
defined with respect to a set of threshold levels $t_k, \,
k=1,\ldots,N_{c}+1$. If $z_{\min} = \min(z_{1},\ldots,z_{N})$,
$z_{\max} = \max(z_{1},\ldots,z_{N}),$ $\delta
t=(z_{\max}-z_{\min})/N_{c}$, the thresholds are defined as: $t_{1}
= -\infty$, $t_2=z_{\min}+\delta t$, $t_{i}=t_{i-1}+\delta t$
$(i=3,\ldots,N_{c})$ and $t_{N_{c}+1}=\infty$.  Each class
${\mathcal C}_q$ corresponds to the interval  ${\mathcal C}_q=(
t_{q}, t_{q+1}]$ for $q=1,\ldots,N_c$.  This means that all the
classes, except the first and the last, have a uniform width. The
classes  ${\mathcal C}_1$ and ${\mathcal C}_{N_c}$ extend to
infinity (negative and positive respectively), to allow for values
at the prediction points that lie outside the observed interval
$[z_{\min}, z_{\max}]$. We define the \textit{class indicator field}
(phase field) $I_{Z}(\vec{r})$ by means of
\begin{equation}
\label{eq:indic}
 I_{Z}(\vec{r}_{i})= \sum_{q=1}^{N_c} q \, \left[ \theta\left( z_{i} - t_{q} \right)
 - \theta\left( z_{i} - t_{q+1} \right) \right], \quad \forall
 i=1,\ldots,N,
\end{equation}
where $\theta(x)=1$ for $x>0$ and  $\theta(x)=0$ for $x \le 0$ is
the unit step function.
 Prediction of the field values $Z\{G_{p}\}$ is
mapped onto a classification problem, i.e., into estimating
$\hat{I}_{Z}\{G_{p}\}$.  If the number of levels is high, the
continuum limit is approached.

\subsection{$k$-nearest neighbor ($k$-NN) classifier} \label{knn_model}

The $k$-NN method is a supervised learning algorithm, which assigns
to each prediction point the class represented by the majority of
its $k$ nearest neighbors from the sample \cite{dasa91}. The
distance metric used is typically Euclidean. The optimal value of
the parameter $k$ depends on the data and requires tuning for
different applications. The optimization of $k$ usually involves
heuristic techniques (e.g. cross-validation). The accuracy of the
$k$-NN classifier is affected by noise or by selecting a
neighborhood that does not include the pertinent spatial
information. The $k$-NN algorithm has been widely applied in the
field of data mining, statistical pattern recognition, image
processing and many others. In general, it has been found to
outperform many other flexible nonlinear methods, particularly in
high-dimensional spaces \cite{cher96}. Furthermore, it is relatively
easy to implement and has desirable consistency properties, e.g.,
favorable asymptotic classification error dependence.


We use the $k$-NN classifier as a benchmark for the ``spin-based''
classification methods we propose. To eliminate the effect of $k$ on
the classification results, for each simulated realization we
perform the classification for a wide range of values of
$k=1,\ldots,k_{\max}$, and select the value $k_{\rm opt}$ that
minimizes  the misclassification rate.

\section{Spatial Classification based on
``spin'' correlation models} \label{models_nnc} We propose two
non-parametric, nearest-neighbor (NN), multilevel correlation (MLC)
models that are loosely based on the Ising and Potts models. The
NN-MLC models are based on matching suitably normalized correlation
energy functions calculated from the samples with those estimated
over the entire prediction grid. Similar ideas of correlation energy
matching were also applied in the reconstruction of digitized random
media from limited morphological information
\cite{yeong98a,yeong98b}. In contrast with those studies, in the
NN-MLC models we also use local spatial information (i.e., the
values at the sampling points). The prediction of the field values
(more precisely, class) at unsampled locations is achieved by means
of conditional simulations that respect locally the sample values
and the correlation energy globally.

\subsection{Ising model}

The Ising model (e.g. \cite{mccoy73}) involves discrete  variables
$s_{i}$ (spins) placed on a sampling grid. Each spin can take two
values, $\pm 1$, and  the spins interact in pairs. Assuming only
nearest-neighbor interactions, the energy of the system can be
expressed by the following Hamiltonian:
\begin{equation*}
\label{Ising} H_{I} = -  \sum_{i,j} J_{ij}s_{i}s_{j}-
\sum_{i}h_{i}s_{i}.
\end{equation*}
The coupling strength $J_{ij}$ controls the type  (ferromagnetic
 for $J_{ij}>0$,  antiferromagnetic for $J_{ij}<0$) and strength of the
 interactions. The second term introduces a
 symmetry-breaking bias caused by the presence of a site
dependent external field $h_{i}$. The latter controls the mean spin
value (the magnetization). The model is usually defined on a regular
grid, the interactions are considered uniform and their range
limited to nearest neighbors. However, the model can be generalized
to include also irregular grids and longer-range interactions.

\subsection{Potts model}

The Potts model is a generalization of the Ising model  (e.g.
\cite{wu82}). Instead of $\pm 1$, each spin is assigned an integer
value $s_{i} \in \{1,\ldots,N_c\}$, where $N_c$ represents the total
number of states or classes. The Hamiltonian of the Potts model is
given by
\begin{equation*}
\label{Potts} H_{P} = - \sum_{i,j}J_{ij}\delta_{(s_{i},s_{j})} -
\sum_{i}h_{i}s_{i},
\end{equation*}
where $\delta$ is the Kronecker delta. Hence, only pairs of spins in
the same state give a non-zero contribution to the correlation
energy. For $N_c=2$, the Potts model is equivalent to the 2D Ising
model.

\subsection{General Assumptions for NN-MLC Classifiers}

The states (configurations) of both the Ising and Potts models are
determined by the Gibbs probability density function $f =
Z^{-1}\,\exp(-H/k_{B}T),$ where $H$ is the respective Hamiltonian,
$k_{B}$ is the Boltzmann's constant, and $T$ is the temperature. The
partition function $Z$ is obtained by summing $e^{-H/k_{B}T}$ over
all possible spin configurations. Essentially, the Hamiltonian
involves two parameters, i.e. the normalized interaction couplings
$\tilde{J}_{ij}=J_{ij}/k_{B}T$ and $\tilde{h}_{i}=h_{i}/k_{B}T.$ In
the case of spatial classification, the parameters are not known
{\em a priori} and need to be determined from the sample.  The
standard procedure for this
 inverse problem  entails using the maximum likelihood method.
 Assuming that the parameters can be inferred,  the
spin values at unsampled locations are predicted by maximizing the
conditional probability density function $f\left(I_{Z}\{G_{p}\}|
I_{Z}\{ G_{s}\};\tilde{J}_{ij},\tilde{h}_{i}\right)$. However,
optimizing the likelihood of the models with respect to the coupling
parameters $\tilde{J}_{ij},\tilde{h}_{i}$ is a computationally
intensive task, since there are no generally valid closed-form
expressions for the partition function. In order to overcome this
problem  we opted for a non-parametric approach.

Our NN-MLC models retain only the interaction energies of the Ising
and Potts spin Hamiltonians. The sample values $Z\{G_s\}$ are mapped
into spin values by suitable discretization (as shown below). The
main idea in both methods is to match the sample correlation energy
 with the correlation energy of all the spins (i.e.
including the simulated spins at prediction points). This relies on
the ergodic assumption that the sample spin correlation energies
accurately describe those of the entire system. The matching of the
correlation energies is performed by means of a numerical Monte
Carlo approach. During the process, the spins at the sample sites
remain fixed. Hence, the procedure followed is conditional
simulation, as opposed to deterministic prediction. Focusing on
regular grids  and assuming isotropic and nearest-neighbor ``spin
interactions,'' it is reasonable to set $J_{ij} = J_{0}$ if the
points $\vec{r}_{i}$, $\vec{r}_{j}$ are lattice neighbors and
$J_{ij} =0$ otherwise. Furthermore, we set $h_{i}=0 \,
(i=1,\ldots,N)$. This choice prevents explicit control of the mean
spin. Nevertheless,  as shown by the simulation results, judicious
choice of the initial state allows the distribution of the predicted
classes to accurately recover the class distribution of the sample.
This is due to the fact that both the correlation energies and the
local spin values are restricted by the sample.

\subsection{Ising-based NN-MLC model (I-NN-MLC)}

We propose a sequential scheme in which the sample, $G_{s}^{q}$ and
prediction, $G_{p}^{q}$ grids are sequentially updated. Each
simulation level $q$ corresponds to a class index. The number of
simulation levels coincides with that of discretization levels. For
the lowest level $G_{s}^{1}=G_{s}, \, G_{p}^{1}=G_{p}.$ For
$q=1,\ldots,N_{c}-1$ it holds that $G_{s}^{q} \subseteq
G_{s}^{q+1}$, $G_{p}^{q} \supseteq G_{p}^{q+1}$ and
$\tilde{G}=G_{s}^{q} \cup G_{p}^{q}.$ Binary-valued spins are used
at each level $q (q=1,\ldots,N_{c})$. Sites (either from the sample
or simulated) having $I_{Z} \le q$ are assigned a spin value of
$-1,$ while sites having $I_{Z}
>q$ are assigned a spin value of $1$. All the sites that are
assigned a spin $-1$ value at level $q$ retain this value at higher
levels. This means that the areas of low values are classified
first. Once a site $\vec{r}_{i}$ is assigned a spin $-1$ value at
level $q<N_c$, it is also assigned class value
$I_{Z}(\vec{r}_{i})=q.$ At the same time, the set $G_{s}^{q+1}$
acquires all the $-1$ points while the set $G_{p}^{q+1}$ is
accordingly reduced. In contrast, for sites that are assigned spin
$1$ value $I_{Z}(\vec{r}_{i})>q$, and the precise class value
$I_{Z}(\vec{r}_{i})$ is determined at a higher level. At level
$q=N_c$ all remaining sites are assigned to the $N_c$-th class.

Let $S^{q}_{s}=\{s_{i}^{q}; \, \forall \, i \, \mbox{s.t.} \,
\vec{r}_{i} \in  G_{s}^{q}\}$, $\forall q=1,\ldots, N_c,$ be the set
of spin values included in the ``sample'' at level $q$. These values
are $\pm 1$ (depending on $z_{i}$) if $\vec{r}_{i} \in  G_{s}$ and
$-1$ if $\vec{r}_{i}$ is an already classified prediction point. The
unknown values at level $q$ are the spins at the remaining
locations, denoted by $S^{q}_{p}$.

Our non-parametric approach utilizes a cost function, $U_{\rm
I}(S^{q}_{p}|S^{q}_{s}),$ that describes the deviation between the
normalized correlation energies of the simulated spin configuration,
$\tilde{C}_{\rm I}^{q}$, and its sample counterpart, $C_{{\rm
I};s}^{q}$ estimated from the spins in $G_{s}^{q}$.
\begin{eqnarray}
\label{eq:cost-i} U_{\rm I}(S^{q}_{p}|S^{q}_{s}) = \left\{
\begin{array}{ll}
\Big[1  - \tilde{C}^{q}_{\rm
I} (S^{q}_{p},S^{q}_{s})/C_{{\rm I};s}^{q}(S^{q}_{s})\Big]^2,
      & {\rm{for}}\ C_{{\rm I};s}^{q} \neq 0, \\ \\
\tilde{C}^{q}_{\rm I}(S^{q}_{p},S^{q}_{s})^2,
& {\rm{for}}\ C_{{\rm I};s}^{q} = 0,
\end{array} \right.
\end{eqnarray}
\noindent where $C_{{\rm I}; s}^{q}(S^{q}_{s})=\langle
s^{q}_{i}\,s^{q}_{j} \rangle _{G_{s}^{q}}$ is the spatially averaged
(normalized by the number of nearest neighbor pairs in $G_{s}^{q}$)
spin pair correlation of the $q-$level sample and
$\tilde{C}^{q}_{\rm I}(S^{q}_{p},S^{q}_{s})=\langle
s^{q}_{i}s^{q}_{j} \rangle _{\tilde{G}}$ is the spatially averaged
spin pair correlation over the entire grid. Given the above, the
estimation of $S^{q}_{p}$ is equivalent to finding the optimal
configuration $\hat{S}^{q}_{p}$ that corresponds to the minimum of
the cost function (\ref{eq:cost-i}) at a fixed temperature $T$,
i.e.,
\begin{eqnarray}
\label{eq:optim-i} \hat{S}^{q}_{p} = {\argmin}_{S^{q}_{p}} \, U_{\rm
I}(S^{q}_{p}|S^{q}_{s}), \, \mbox{for} \; q=1,\ldots,N_c.
\end{eqnarray}
%

\subsection{Potts-based NN-MLC model (P-NN-MLC)}
In the P-NN-MLC model the classification is performed for all
classes simultaneously. Hence, there is only a single simulation
level irrespectively of the number of discretization levels. The
grid spins $S=\{s_{i}\}$, $i=1,\ldots,N$ take values in the set $s_i
\in \{1,\ldots,N_c\}$. The cost function is given by
\begin{eqnarray}
\label{eq:cost-p} U_{\rm P}(S_{p}|S_{s}) = \left\{
\begin{array}{ll}
\Big[1  - \tilde{C}_{\rm P}(S_{p},S_{s})/C_{{\rm P};s}(S_{s})\Big]^2,
  & {\rm{for}}\ C_{{\rm P};s} \neq 0, \\ \\
\tilde{C}_{\rm P}(S_{p},S_{s})^2,                             &
{\rm{for}}\ C_{{\rm P};s} = 0,
\end{array} \right.
\end{eqnarray}
\noindent where $C_{{\rm P};s}(S_{s})=\langle \delta_{(s_{i},s_{j})}
\rangle_{G_{s}}$ is the spatially averaged (normalized by the number
of nearest neighbor pairs in $G_{s}$) spin pair correlation of the
sample and $\tilde{C}_{\rm P}(S_{p},S_{s})=\langle
\delta_{(s_{i},s_{j})} \rangle _{\tilde{G}}$ is the average spin
pair correlation over ${\tilde{G}}$. The estimation of $S_{p}$ is
equivalent to finding the minimum of the cost function
(\ref{eq:cost-p}), i.e.,
\begin{eqnarray}
\label{eq:optim-p} \hat{S}_{p} = {\argmin}_{S_{p}} \, U_{\rm
P}(S_{p}|S_{s}).
\end{eqnarray}
%
%

\section{Simulations of Missing Data on Regular Grids}
\label{simul} We focus on samples with missing data on regular
grids. On such grids, it is straightforward to determine the nearest
neighbors and calculate the correlation energies. Both the I-NN-MLC
and P-NN-MLC methods return a class indicator field [see
Eq.~(\ref{eq:indic})] $\hat{I}_{Z}=I_{Z}(G_{s})\cup
\hat{I}_{Z}(G_{p})$, which consists of the original sample
classification and the class estimates at $G_{p}$. The indicator
values at the sampling sites are exactly reproduced. Below we refer
to $I_{Z}(G_{s})$ as the training set. Optimization of the cost
functions~(\ref{eq:cost-i}) and (\ref{eq:cost-p}) are performed
using the Monte Carlo approach.  The generation of new ``trial''
spin states  is realized using the Metropolis algorithm at zero
temperature.

We use  the {\em rejection ratio} defined by $\rho=
(\#\mbox{rejected states})/N(G_{p}^{q}),$ where $N(G_{p}^{q})$ is
the number of prediction points at the $q$-level, to determine the
stopping criterion. More specifically, our simulations terminate if
$\rho=1$, i.e., if one complete sweep through the entire grid
$G_{p}^{q}$ does not produce a single successful update. Reaching
the termination criterion may require several sweeps through the
lattice, depending on the initial state.

\subsection{Greedy Monte Carlo}

 The $T=0$ assumption implies that the stochastic
selection of an energetically unfavorable spin configuration in the
Metropolis step is not possible. Hence, the candidate ``spin'' for
the update is flipped unconditionally only if it lowers the cost
function. This approach is called the ``greedy'' Monte Carlo
algorithm~\cite{pap82} and leads to very fast convergence. In
contrast, in simulated annealing $T$ is slowly lowered starting from
an initial high-temperature state. This approach is much slower
computationally. However, the configuration resulting from simulated
annealing is less sensitive to the initial state. The sensitivity of
the greedy algorithm is known to be especially pronounced in
high-dimensional spaces with non-convex energies. In such cases, the
greedy algorithm is likely to be trapped in local minima, instead of
converging to the global one. However, this is not a concern in the
current problem. In fact, targeting the global minimum of $U_{\rm
I}$ and $U_{\rm P}$ strongly emphasizes the sample correlation
energy per ``spin'' pair. However, the latter is influenced by
sample-to-sample fluctuations.

On rectangular or square grids, further increase in computational
efficiency is gained by taking advantage of the geometry and the
nearest-neighbor interactions. This is achieved by splitting the
grid into two interpenetrating subgrids, which allows vectorizing
the algorithm. Hence, one sweep through the entire grid is performed
in just two steps by simultaneously updating all the sites on one of
the subgrids in each step.

\subsection{Simulation Algorithms}
Based on the above, the monte Carlo (MC) algorithms for the I-NN-MLC
and P-NN-MLC models consist of the following steps:

\begin{center}\textbf{Algorithm for I-NN-MLC model} \end{center}
 \noindent {\bf (1)} {\em Initialize the
indicator field on the entire grid by means of}
$\hat{I}_{Z}(\tilde{G})={\rm NaN}$
\newline
{\bf (2)} {\em Set the simulation level (class index) to $q=1$}
\newline
{\bf (3)} {\bf While [loop 1]} $q \le N_c -1$ {\em discretize}
$Z\{G_{s}\}$ {\em with respect to} $t_{q+1}$ {\em to obtain}
$S^{q}_{s}$
\newline
 \indent {\rm (3.1)} {\em Given the data} $S^{q}_{s}$,
 {\em calculate the sample
correlation energy} $C_{I;s}^{q}$
\newline \indent
{\rm (3.2)} {\em Assign initial values to the spins at} $G^{q}_{p},$
{\em i.e., generate} $\hat{S}^{q \,(0)}_{p}$
\newline
 \indent {\rm (3.3)} {\em Calculate the initial values of the simulated
correlation} $\tilde{C}_{I}^{q \, (0)}$
\newline \indent
{\em and the cost function} $U_{I}^{(0)}$; {\em initialize the MC
index} $i=1$
\newline
 \indent {\rm (3.4)} {\em Initialize the rejection ratio} $\rho \rightarrow
0;$
 {\em and the rejected states
index} $i_r =0$
\newline \indent
 {\rm (3.5)} {\bf While [loop 2]} $\rho < 1$ {\em repeat the
following updating steps}:
\newline \indent \indent {\rm (3.5.1)}{\em Generate a new state}
$\hat{S}^{q \, (i+1)}_{p}$ {\em by perturbing} $\hat{S}^{q \,
(i)}_{p}$
\newline \indent \indent {\rm (3.5.2)}{Calculate} $\tilde{C}_{I}^{q \,
(i+1)}$ {\em and} $U_{I}^{(i+1)}$
\newline \indent \indent {\rm (3.5.3)} {\em If} $U_{I}^{(i+1)} < U_{I}^{(i)}$
{\em accept the new state;} $i_r \rightarrow 0.$
\newline \indent \indent \indent {\em else
 keep the ``old'' state}; $i_r \rightarrow i_r+1$; {\em endif}
 \newline \indent  \indent {\rm (3.5.4)}
 $\rho \rightarrow i_r/N(G_{p}^{q}); \, i \rightarrow i+1$;
\newline \indent \indent   {\bf end [loop 2]}
\newline \indent {\rm (3.6)} {\em Assign the}  $-1$
{\em ``spins" to the $q$ level,  i.e.},
$\hat{I}_{Z}(\{\vec{r}_{i}\})=q$ \newline \indent \indent {\em If}
$\hat{S}^{(i_{\max})}(\{\vec{r}_{i}\})=-1$, $\{\vec{r}_{i}\} \in
\tilde{G}$
\newline \indent
{\rm (3.7)} {\em Increase simulation level,} $q \rightarrow q+1$,
{\em return to step {\bf (3)}
\newline \indent {\bf end [loop 1]}
\newline
{\bf (4)} {\em For $q=N_c$,} $\forall \vec{r}_{i} \;
(i=1,\ldots,N_{\tilde{G}})$ {\em such that}
$\hat{I}_{Z}(\{\vec{r}_{i}\})={\rm NaN}$, set
$\hat{I}_{Z}(\{\vec{r}_{i}\})=N_c \; $}.

\vspace{12pt} In the above, the symbol NaN is used to denote
non-numeric values.

\begin{center}\textbf{Algorithm for P-NN-MLC model} \end{center}

 \noindent {\bf (1)}
{\em Discretize} $Z\{G_{s}\}$ {\em with respect to} $t_{k},\
k=1,\ldots,N_{c}+1$ {\em to obtain} $S_{s}$
\newline
 {\bf (2)} {\em Given the data} $S_{s}$, {\em calculate the sample
correlation energy} $C_{P;s}$
\newline
{\bf (3)} {\em Assign initial values to the spins at} $G_{p},$ {\em
i.e., generate} $\hat{S}^{(0)}_{p}$
\newline
{\bf (4)} {\em Calculate the initial values of the simulated
correlation} $\tilde{C}_{P}^{(0)}$ \newline \indent {\em and the
cost function} $U_{P}^{(0)}$; {\em initialize the MC index} $i=1$
\newline
{\bf (5)} {\em Initialize the rejection ratio} $\rho \rightarrow 0;$
 {\em and the rejected states
index} $i_r =0$
\newline
 {\bf (6)} {\bf While} $\rho < 1$ {\em repeat the
following updating steps}:
\newline \indent {\rm (6.1) Generate a new state}
$\hat{S}^{(i+1)}_{p}$ {\em by perturbing} $\hat{S}^{(i)}_{p}$
\newline \indent {\rm (6.2) Calculate} $\tilde{C}_{P}^{(i+1)}$ {\em and} $U_{P}^{(i+1)}$
\newline \indent {\rm (6.3) If} $U_{P}^{(i+1)} < U_{P}^{(i)}$ {\em accept the
new state;} $i_r \rightarrow 0.$
\newline \indent \indent {\em else
 keep the ``old'' state}; $i_r \rightarrow i_r+1$; {\em endif}
 \newline \indent  {\rm (6.4)} $\rho \rightarrow i_r/N(G_p);
 \, i \rightarrow i+1$;
\newline \indent {\bf end}
\newline  {\bf (7)} {\em Assign}
$\hat{I}_{Z}(\{\vec{r}_{i}\})=\hat{S}^{(i_{\max})}(\{\vec{r}_{i}\})$,
$\{\vec{r}_{i}\} \in \tilde{G}$.

\subsection{Initial state selection}

The initial configuration of class indices can be selected in a
number of ways. Since the proposed models aim to provide fast and
automatic classification mechanisms, the initial configuration
(assigned in steps (3.2) and (3) in the I-NN-MLC and P-NN-MLC
algorithms) should minimize the relaxation path in state space to
the equilibrium. It should also be selected with little or no user
intervention. Since a degree of spatial continuity is common in
geospatial data sets, it makes sense that the initial state of the
individual prediction points is determined based on the sample
states in their immediate neighborhood. On square grids, we
determine the neighborhood of ${\vec{r}_p}$ by an $m \times m$
stencil (where $m=2l+1$) centered at ${\vec{r}_p}$. Then, the
initial value at a prediction point is assigned by majority rule,
based on the prevailing value of its sample neighbors  inside the
stencil. If we considered a circular stencil, this method would
correspond to the $k$-NN classification algorithm, $k$ being the
number of sampling points inside the stencil.

\begin{figure}[!t]
  \begin{center}
    \includegraphics[width=0.5\textwidth]{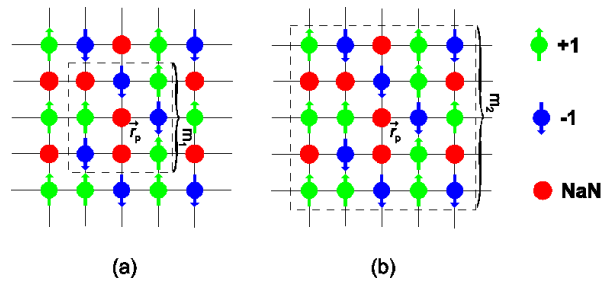}
  \end{center}
  \caption{Schematic demonstrating the stencil selection.
  Arrows pointing up correspond to $+1$ spins,
  arrows pointing down to $-1$ spins, and the empty circles to
  the prediction location.
  In the first plot (a) a square stencil of size $m_1$=3 is used.
  Within the neighborhood marked
  by the dash-line square there is an equal number of $+1$ and $-1$ sites.
  In the second plot (b)
  a stencil of size $m_2=5$ is used to break the tie.}
  \label{fig:stencil_size}
\end{figure}

We set the stencil size adaptively to the smallest size that
contains a clear majority of sample spin values, as shown
schematically in Fig.~(\ref{fig:stencil_size}). In practice, it
makes sense to impose an upper limit on the stencil size $m_{\max}$.
If no majority is established up to the maximum stencil size
$m_{\max}\times m_{\max}$, the initial value at ${\vec{r}_p}$ is
assigned randomly from the equally represented class indices with
the highest frequency (in the I-NN-MLC this means $\pm 1.$) If
majority is not reached due to absence of sampling points inside the
maximum stencil, the initial value is drawn randomly from the entire
range of the labels $1,\ldots,N_c$. There are sensible reasons for
imposing a maximum stencil size. First, considering too large
neighborhoods in the $k$-NN classification generally generates
oversmoothing at larger-scales that can not be justified as an
effect of {\it local} continuity. Second, large neighborhoods
increase the computational demands (both for memory and CPU time)
disproportionately to expected benefits. Finally, assigning a
portion of the prediction points initial values at random introduces
a degree of randomness, which can be used to assess uncertainty by
performing multiple runs on the same sample set. The choice of
$m_{\max}$ is arbitrary to an extent. Intuitively, for sparser
sampling patterns larger $m_{\max}$ should be considered. In our
investigations, relatively small sizes (up to $m_{\max}=7$) were
sufficient to establish good statistical performance at relatively
small computational cost.

The algorithms thus require that only  two parameters be set:
$m_{\max}$ and $N_c$. The latter depends on the study's objectives:
if the goal is to determine exceedance levels of a pollutant
concentration with respect to a regulatory threshold, a binary
classification is adequate. For environmental monitoring and
decision making purposes a moderate number of classes (e.g., eight)
is often sufficient.

\section{Case Study: Missing Data Reconstruction}
\label{results}

\begin{figure}[!t]
  \begin{center}
    \subfigure[Map]{\label{fig:orig_map}\includegraphics[width=0.48\textwidth]{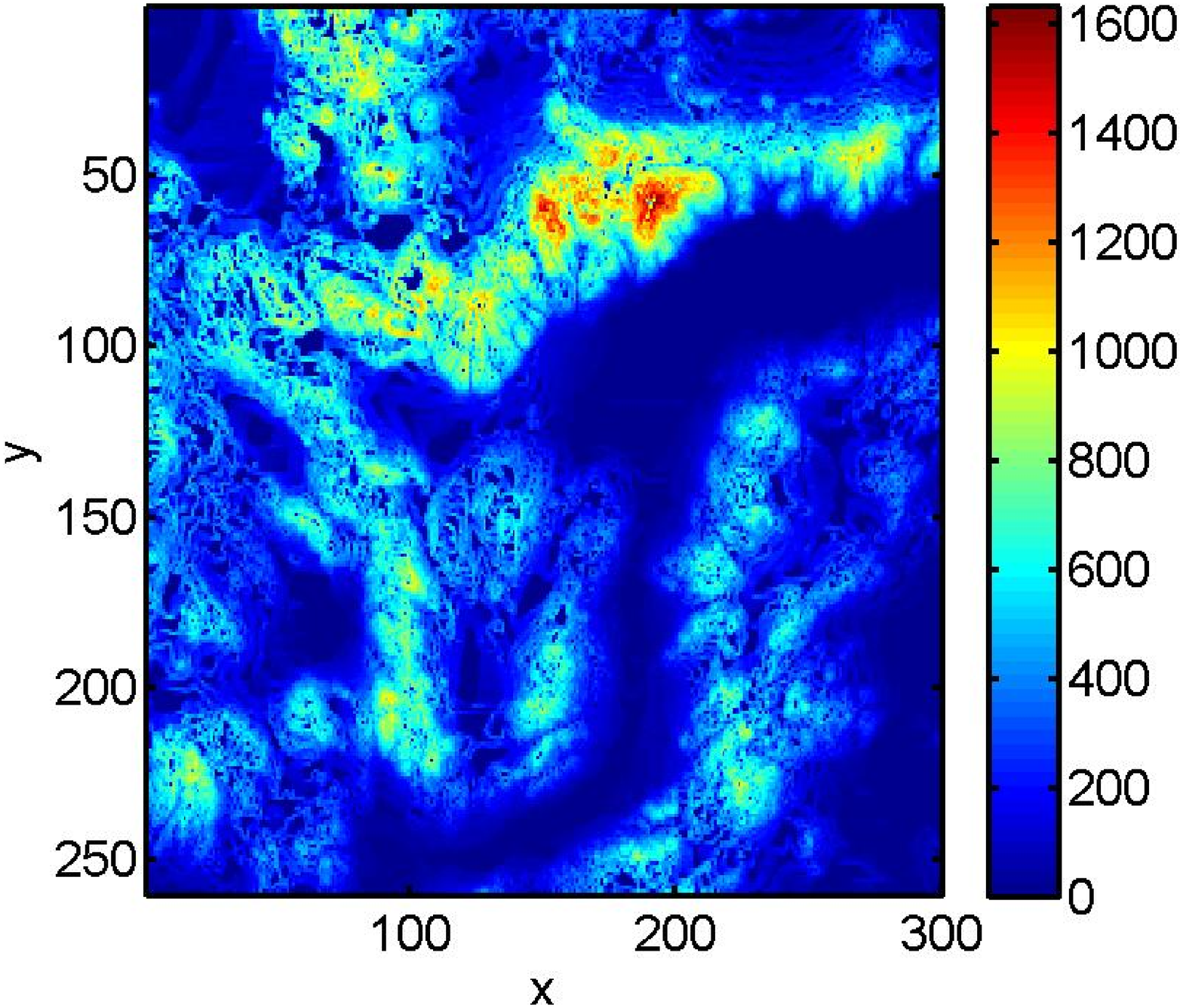}}
    \subfigure[Histogram]{\label{fig:orig_hist}\includegraphics[width=0.45\textwidth]{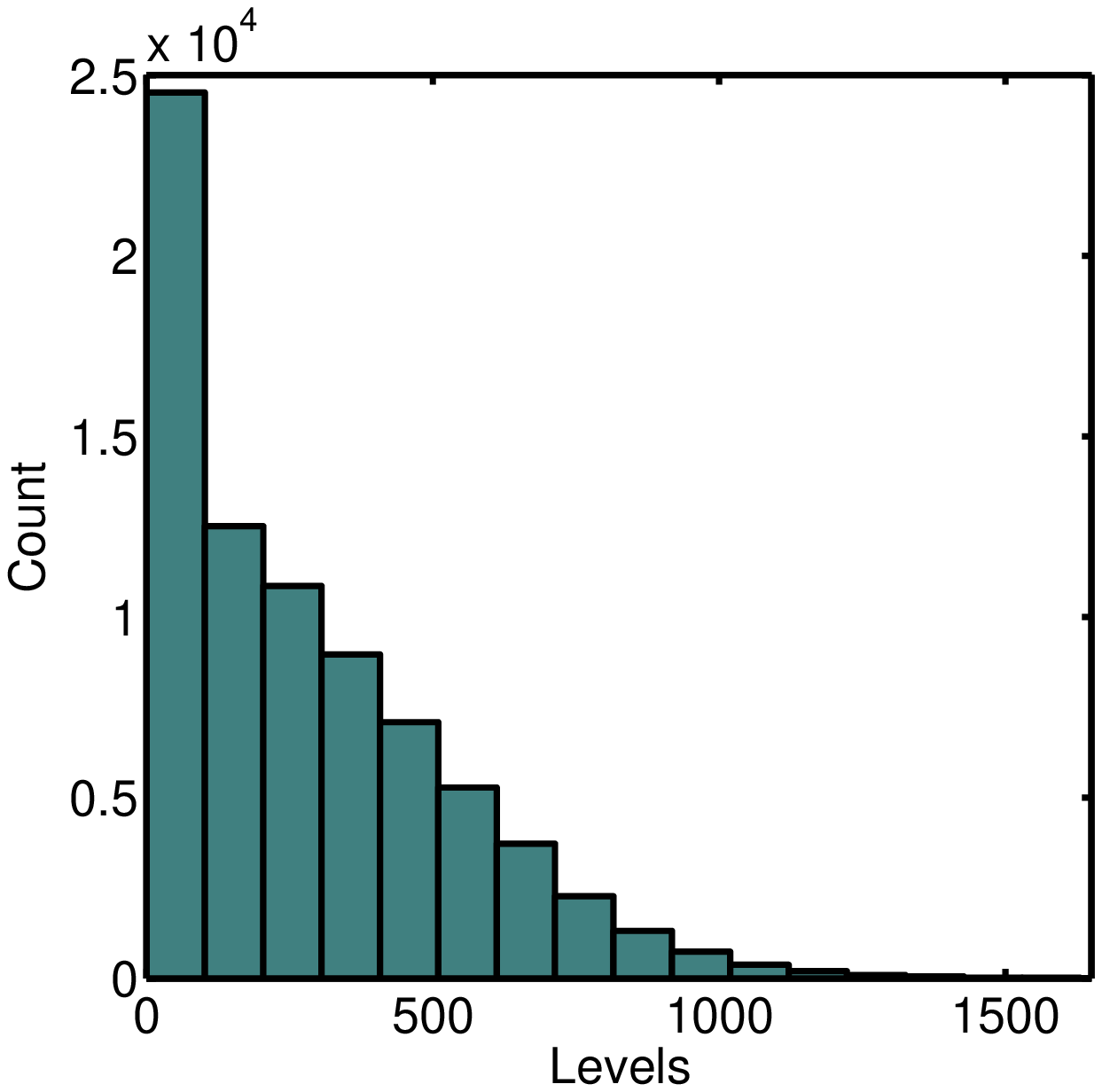}}
  \end{center}
  \caption{Map and histogram of the original complete data.}
  \label{fig:orig}
\end{figure}

To test the classification models we use a synthetic pollutant
concentration data set derived from a digital elevation model of the
Walker lake area in Nevada \cite{isak89}. A two-dimensional
projection of the pollution field is shown in Fig.
\ref{fig:orig_map}. The units used for the $Z$ values are
arbitrarily set to parts per million (ppm). Some summary statistics
are as follows: number $N_{\tilde{G}}=78\ 000$ on a $260 \times 300$
rectangular grid, $z_{\min}= 0$, $z_{\max}= 1631.2$,
$\bar{z}=277.9$, $z_{0.50}= 221.3$, $\sigma_{z}= 249.9$, the
skewness coefficient is $1.02$, and the kurtosis coefficient $3.78$.
As evidenced from the above statistics and the histogram in Fig.
\ref{fig:orig_hist}, the distribution is clearly neither Gaussian
nor log-normal.

\subsection{Computational details}

>From the complete data  we draw a sample $Z\{G_{s}\}$ (training set)
of size $N=(1-p)N_{\tilde{G}}$ by randomly removing $P=p
N_{\tilde{G}}$ values (validation set), which are later used for
prediction validation. For three degrees of thinning, $p=[0.33, 0.50,
0.66]$, we generate $100$ different configurations of the training
and validation sets. The values at the prediction points are
predicted (classified) using the I-NN-MLC and P-NN-MLC
classification algorithms. The class indicator values at the
prediction points $I_{Z}(G_{p})$ are then compared with the
classification estimates $\hat{I}_{Z}(G_{p})$. To evaluate the
classification performance, we calculate the misclassification rate
as a fraction of misclassified pixels:
\begin{eqnarray}
\label{misclass} F = \frac{1}{P}\sum_{p=1}^{P}\left[
1-\delta\big(I_{Z}(\vec{r}_p),\hat{I}_{Z}(\vec{r}_p)\big) \right],
\end{eqnarray}
where $I_{Z}(\vec{r}_p)$ is the true value at the validation points,
$\hat{I}_{Z}(\vec{r}_p)$ is the classification estimate and
$\delta(I,I')=1$ if $I=I'$, $\delta(I,I')=0$ if $I \neq I'$. The standard
deviation of the quantity $F$ is evaluated using the values obtained from
all the configurations, as
${\rm STD}_{F} = \sqrt{\sum_{i=1}^{100}(F_{i}^*-\overline{F_{i}^*})^2/99}$.
We also compare the class index variograms of the original and
reconstructed patterns in the orthogonal lattice directions, defined
by
\begin{equation}\label{vario}
\hat{\gamma}_{\iota}(h)=\frac{1}{2|N_{\iota}(h)|}\sum_{i,j\in N_{\iota}(h)}|I_{Z}(\vec{r}_i)-I_{Z}(\vec{r}_j)|^2,
\end{equation}
where $N_{\iota}(h)$ denotes the set of pairs of points in $\iota$ orthogonal lattice
direction separated by distance $h$ (lag), and $|N_{\iota}(h)|$ denotes the number of pairs
in the set.

Furthermore, we record the CPU time, the number of
MC sweeps needed to reach equilibrium, and the residual values of
the cost functions at termination. The procedure is repeated for
$N_c=8$ and $N_c=16$. The computations are performed in the
Matlab\textregistered\ programming environment on a desktop computer
with 1.93 GB of RAM and an Intel\textregistered Core\texttrademark2
CPU 6320 processor at 1.86 GHz.

%

\begin{figure}[!ht]
    \subfigure[I-NN-MLC, $p=0.33$]{\label{fig:frst-M8-msng33}
    \includegraphics[width=0.48\textwidth]{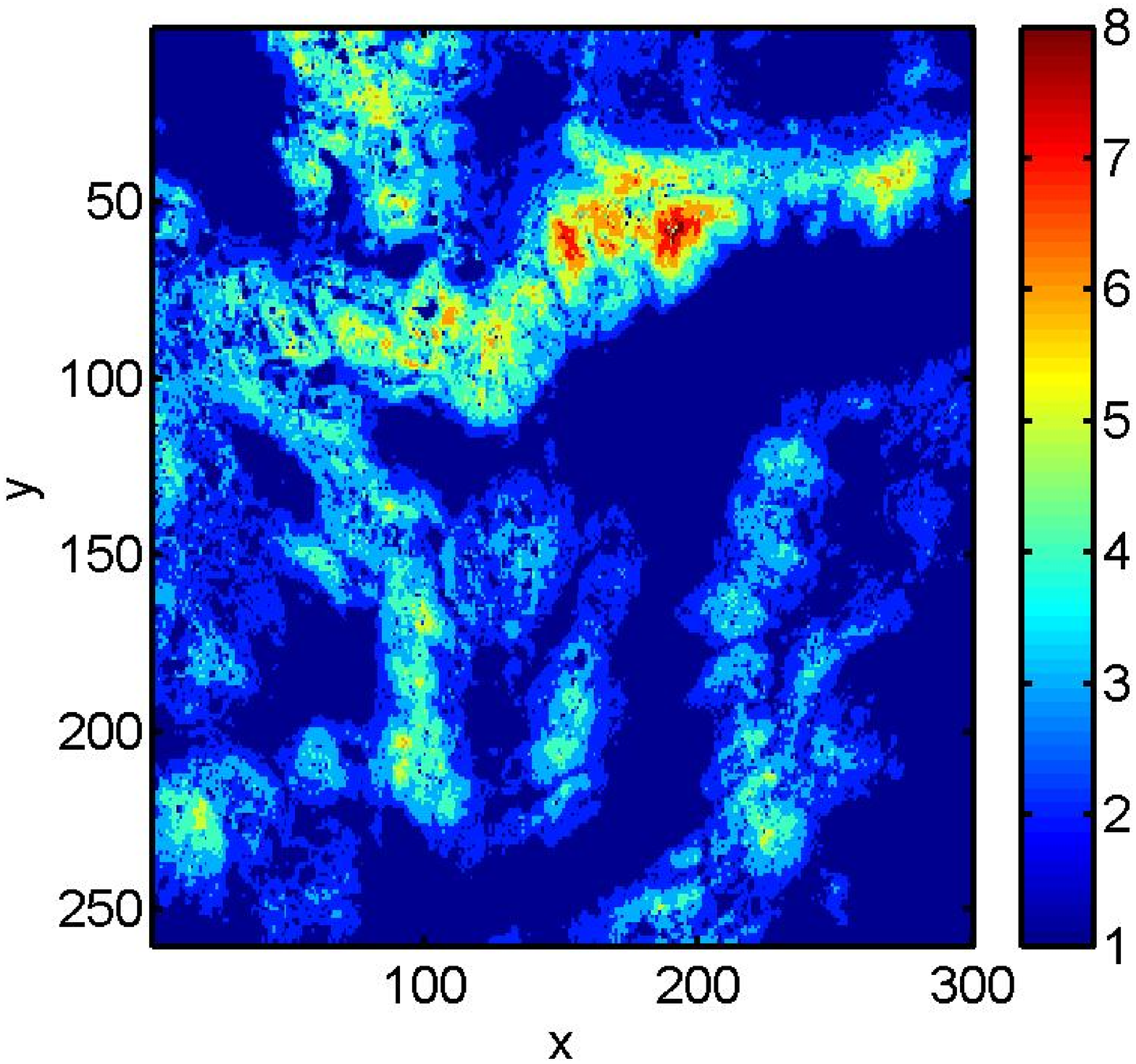}}
    \subfigure[I-NN-MLC, $p=0.66$]{\label{fig:frst-M16-msng66}
    \includegraphics[width=0.48\textwidth]{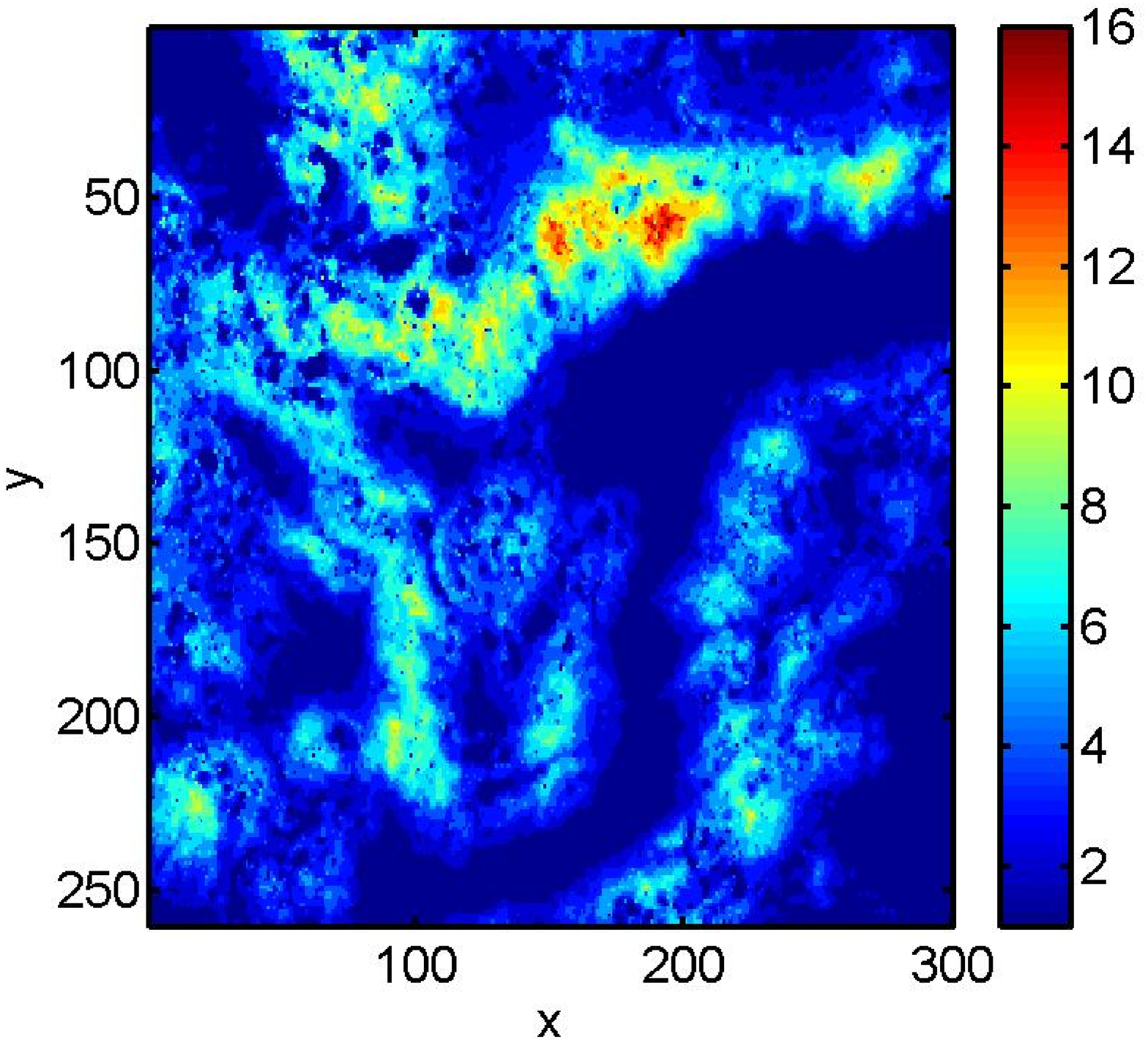}}\\
    \subfigure[P-NN-MLC, $p=0.33$]{\label{fig:frst-q8-msng33}
    \includegraphics[width=0.48\textwidth]{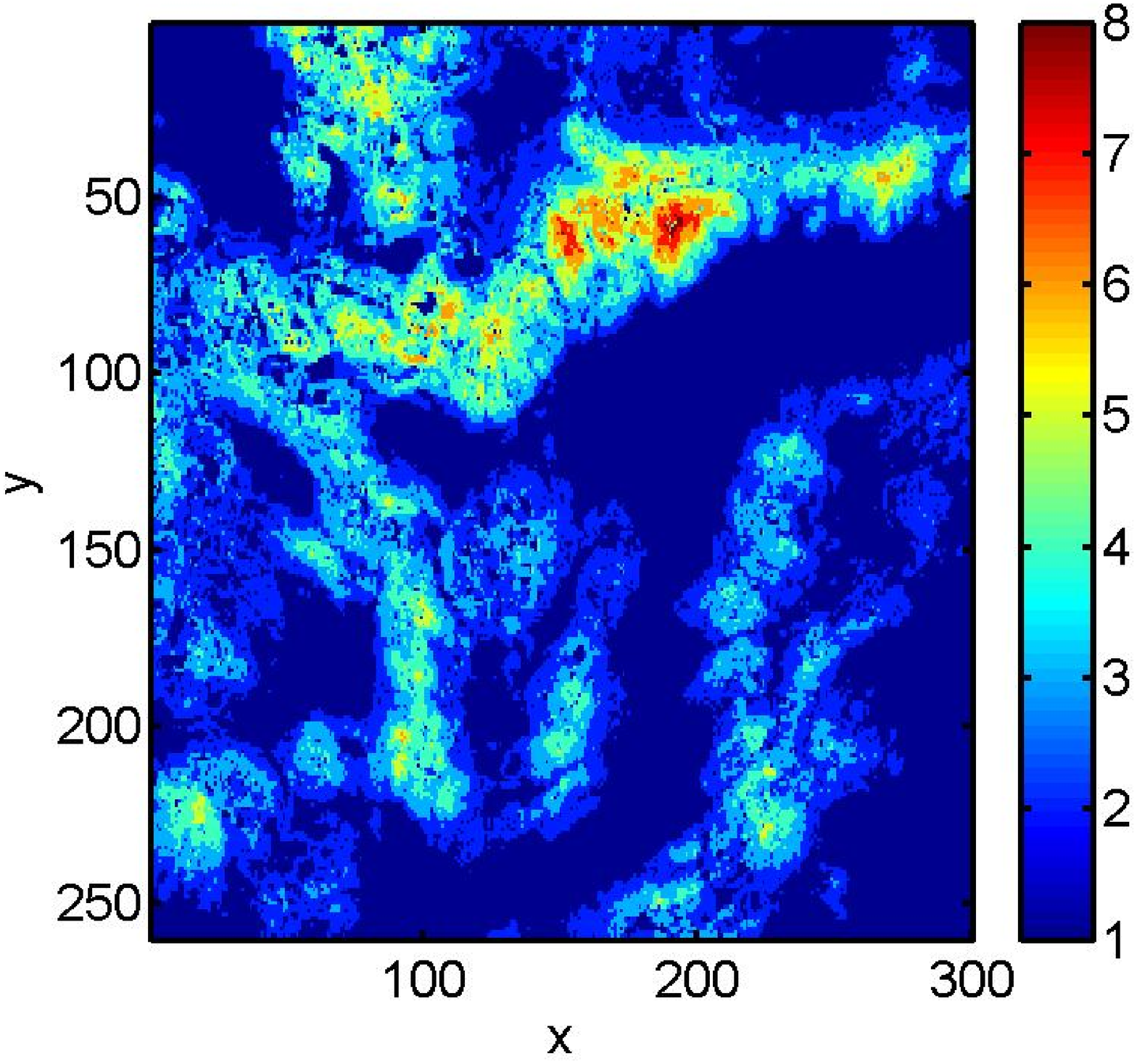}}
    \subfigure[P-NN-MLC, $p=0.66$]{\label{fig:frst-q16-msng66}
    \includegraphics[width=0.48\textwidth]{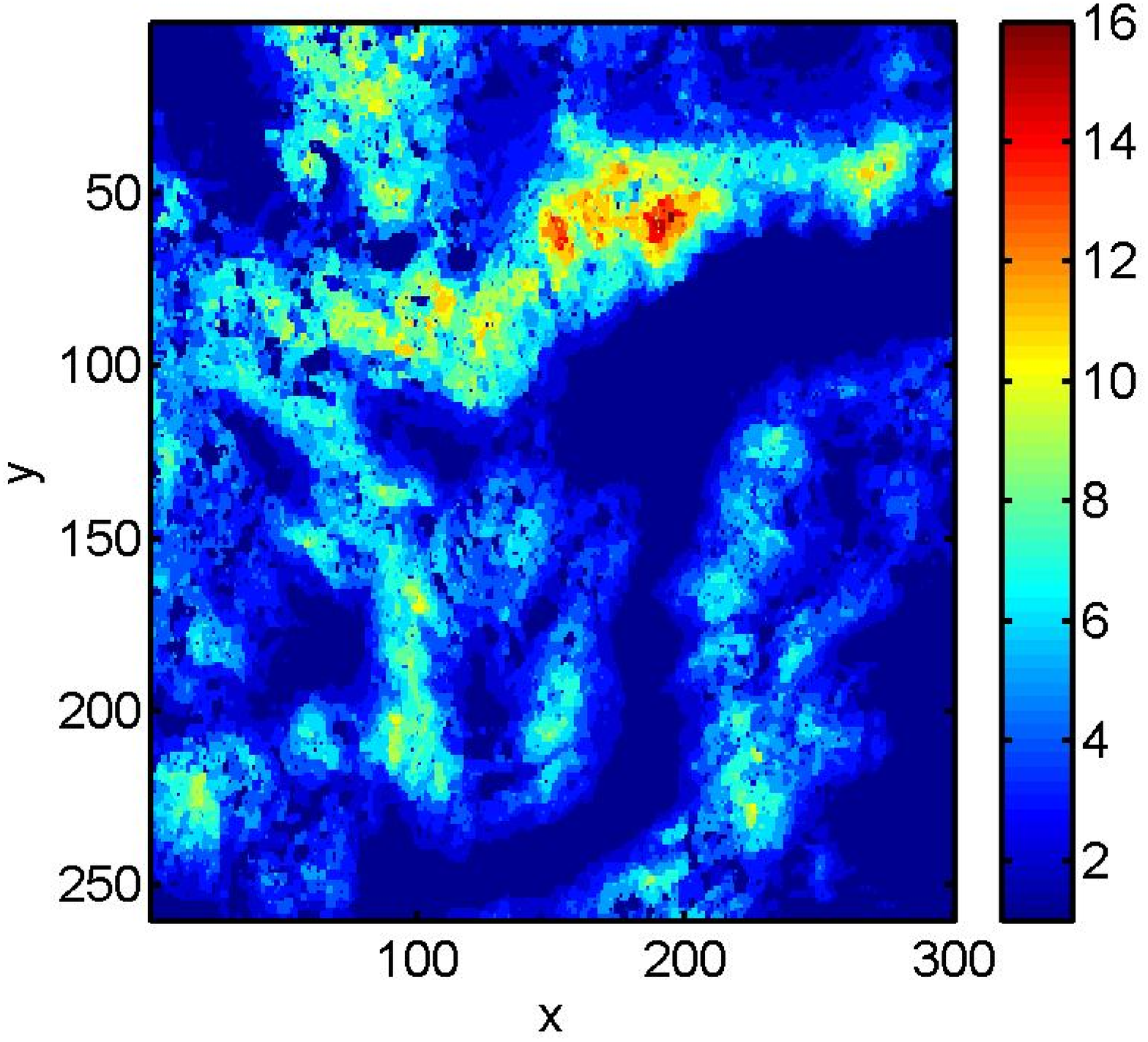}}
  \caption{The reconstructed maps corresponding to the first realization
  obtained by (a) I-NN-MLC with $p=0.33$, (b) I-NN-MLC with $p=0.66$,
  (c) P-NN-MLC with $p=0.33$, and (d) P-NN-MLC with $p=0.66$.}
  \label{fig:rec_maps}
\end{figure}

\begin{figure}[!ht]
    \subfigure[I-NN-MLC, $p=0.33$]{\label{fig:std-M8-msng33}
    \includegraphics[width=0.48\textwidth]{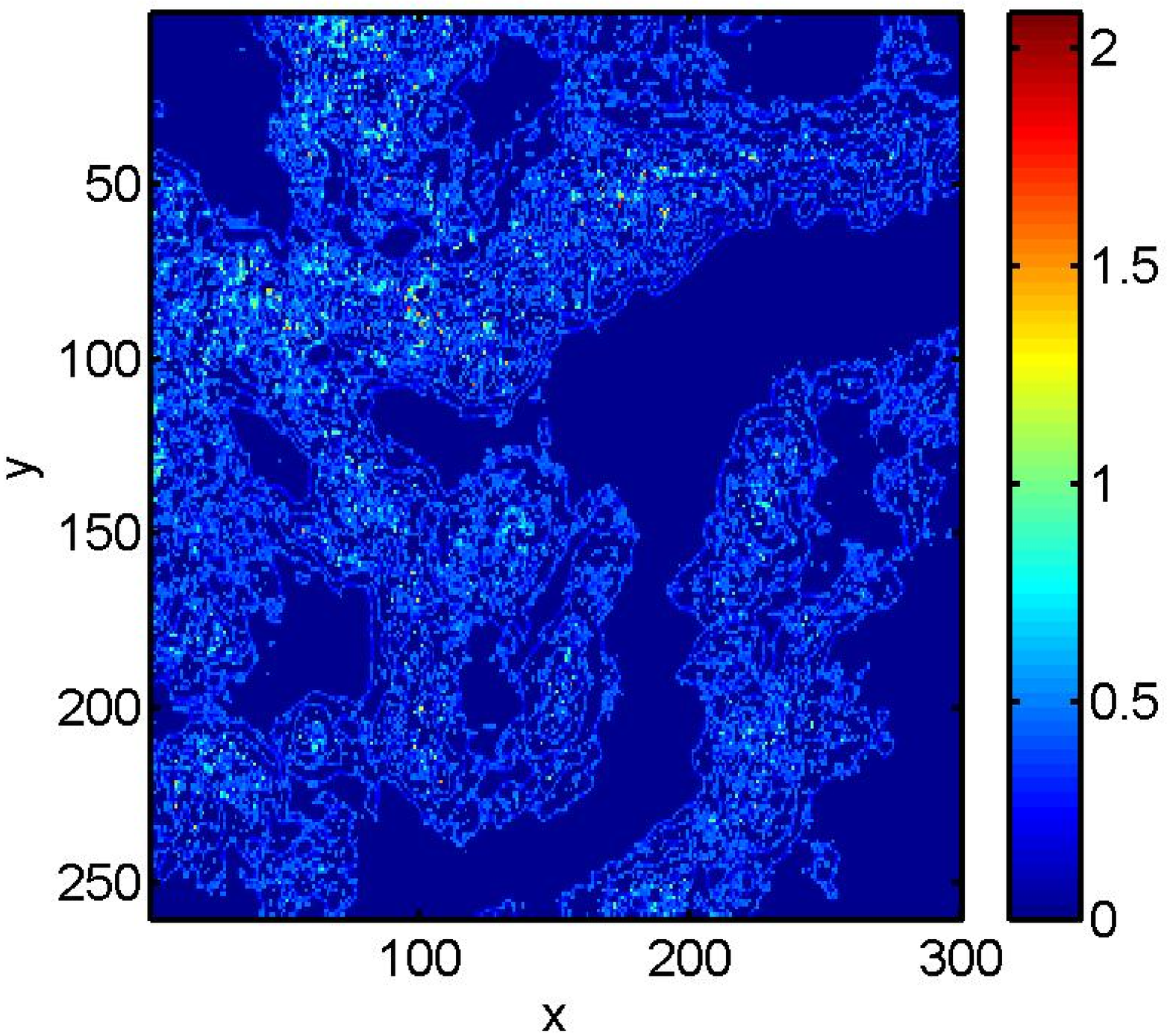}}
    \subfigure[I-NN-MLC, $p=0.66$]{\label{fig:std-M16-msng66}
    \includegraphics[width=0.48\textwidth]{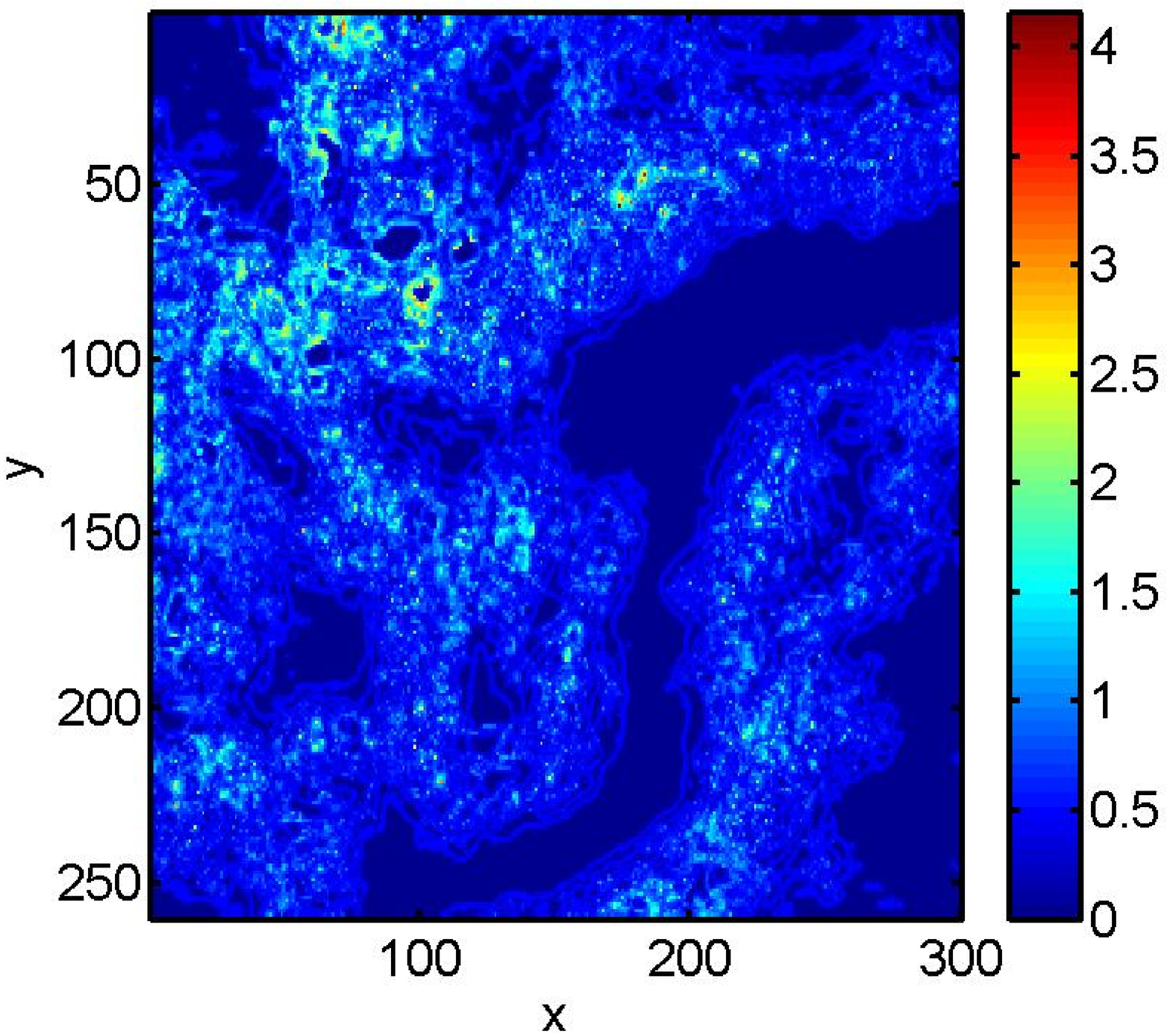}}\\
    \subfigure[P-NN-MLC, $p=0.33$]{\label{fig:std-q8-msng33}
    \includegraphics[width=0.48\textwidth]{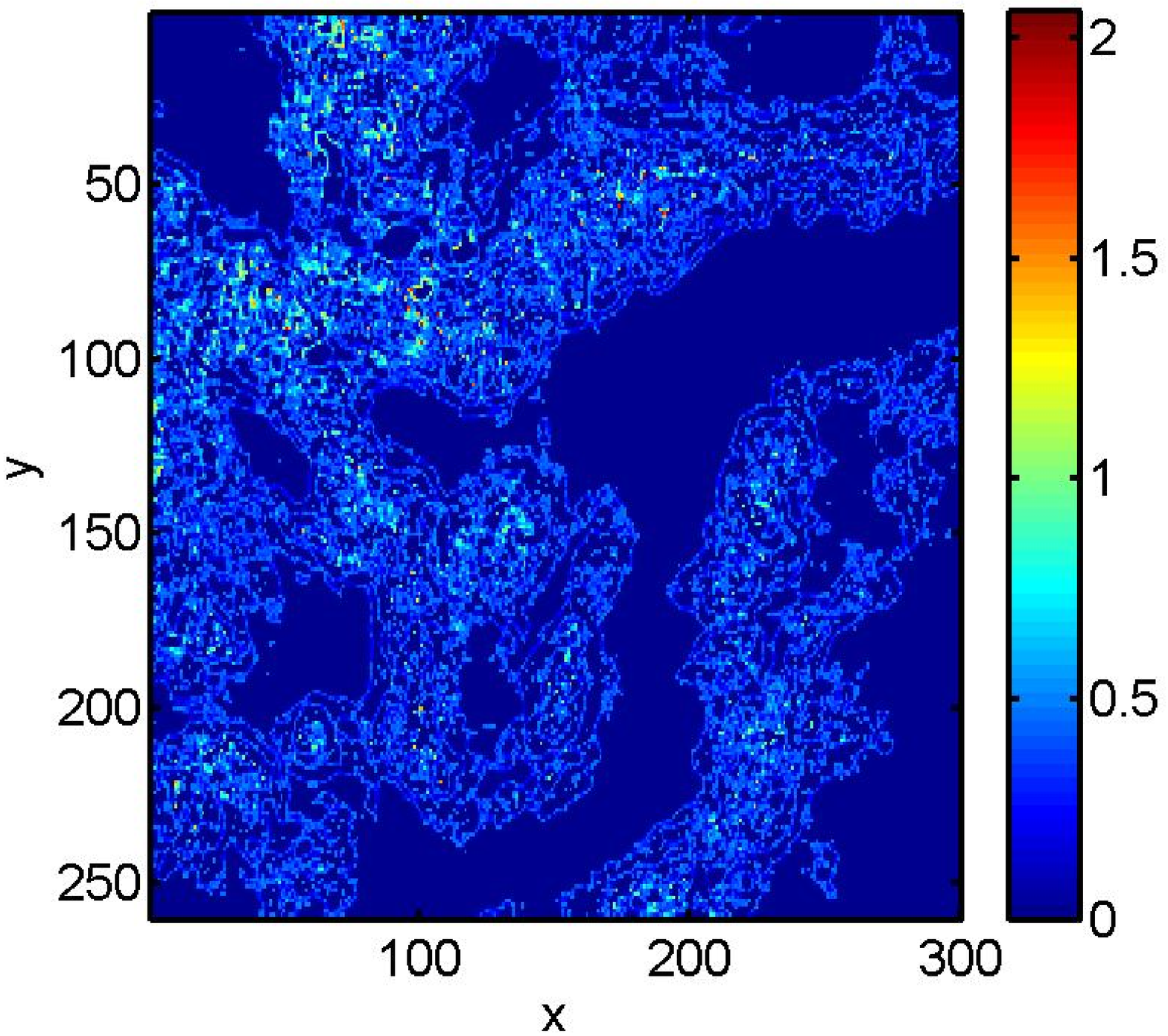}}
    \subfigure[P-NN-MLC, $p=0.66$]{\label{fig:std-q16-msng66}
    \includegraphics[width=0.48\textwidth]{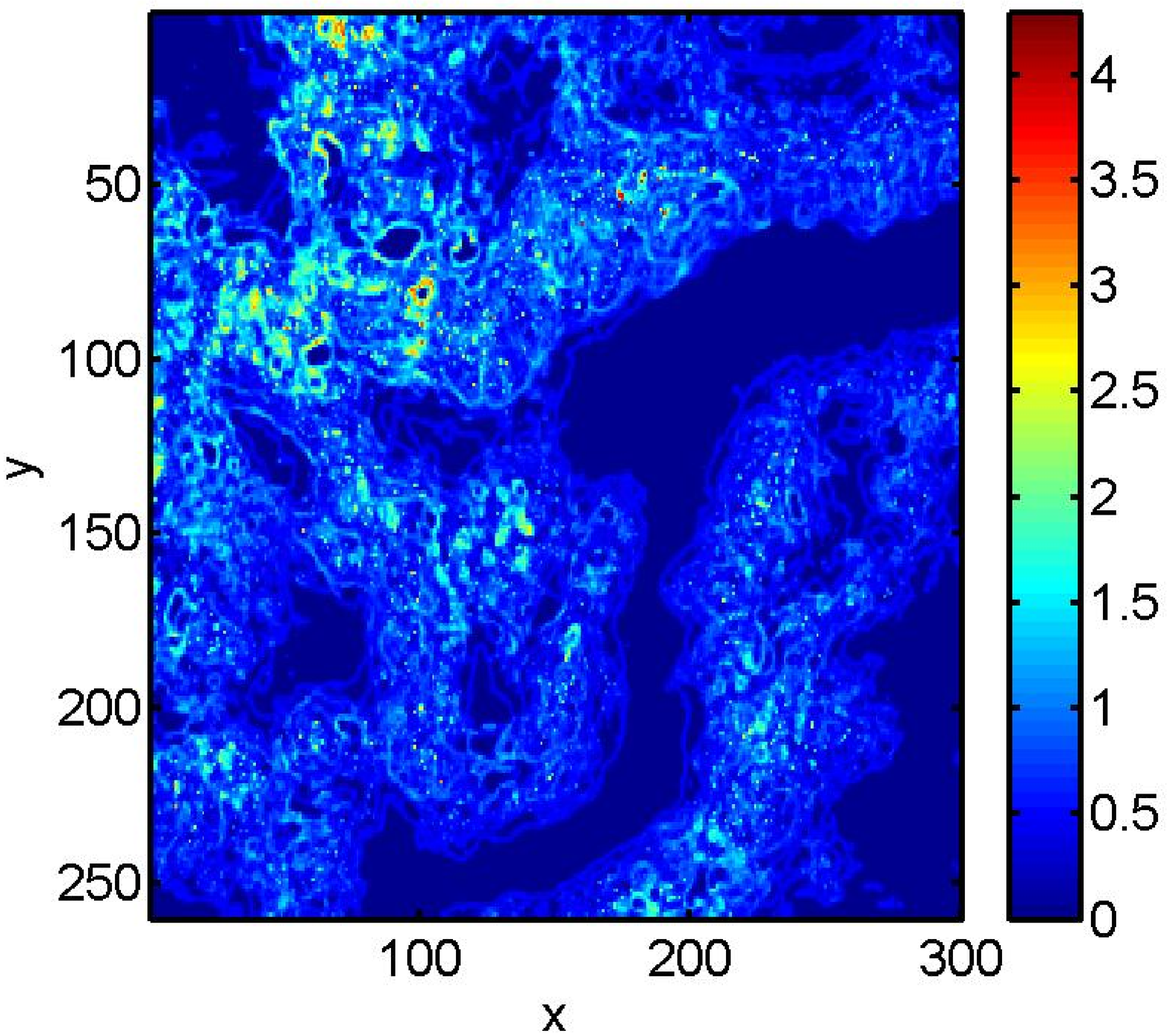}}
  \caption{The standard deviations calculated based on the classification
  results obtained from $100$ realizations obtained by (a) I-NN-MLC with
  $p=0.33$, (b) I-NN-MLC with $p=0.66$, (c) P-NN-MLC with $p=0.33$, and
  (d) P-NN-MLC with $p=0.66$.}
  \label{fig:rec_std}
\end{figure}

\begin{figure}[!ht]
    \subfigure[I-NN-MLC, $p=0.33$]{\label{fig:hist-M8-msng33}
    \includegraphics[scale=0.33]{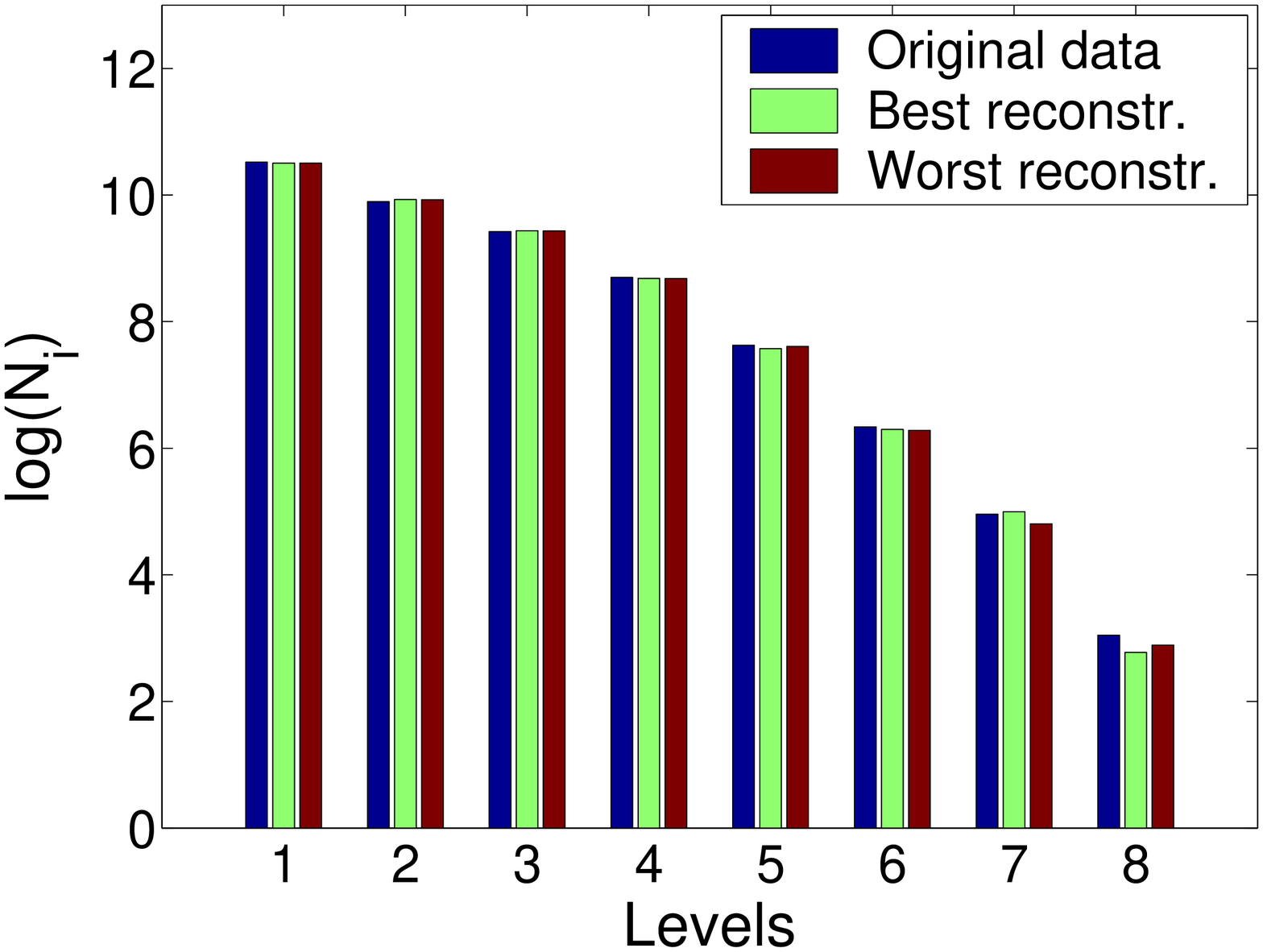}}
    \subfigure[I-NN-MLC, $p=0.66$]{\label{fig:hist-M16-msng66}
    \includegraphics[scale=0.33]{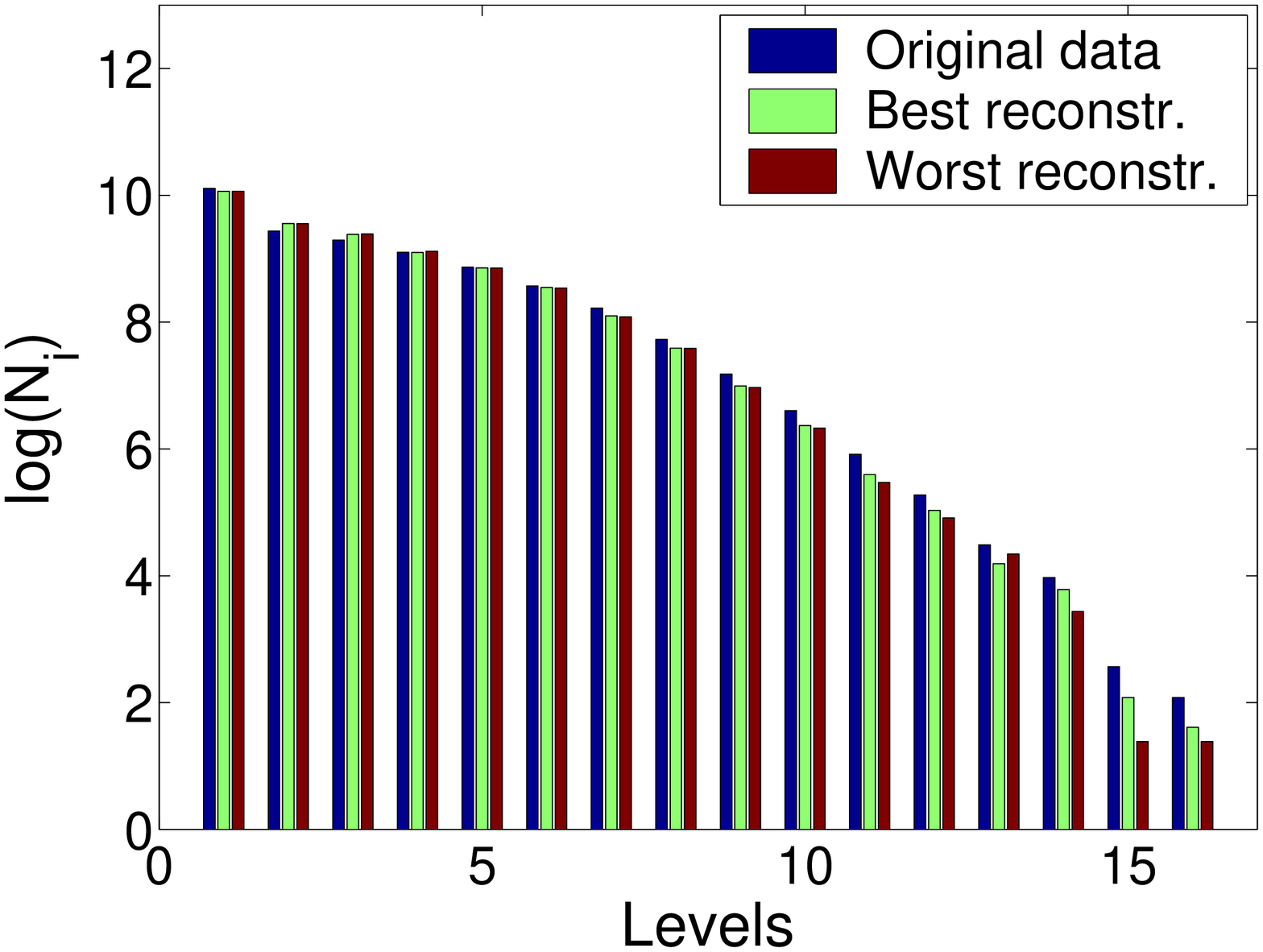}}\\
    \subfigure[P-NN-MLC, $p=0.33$]{\label{fig:hist-q8-msng33}
    \includegraphics[scale=0.33]{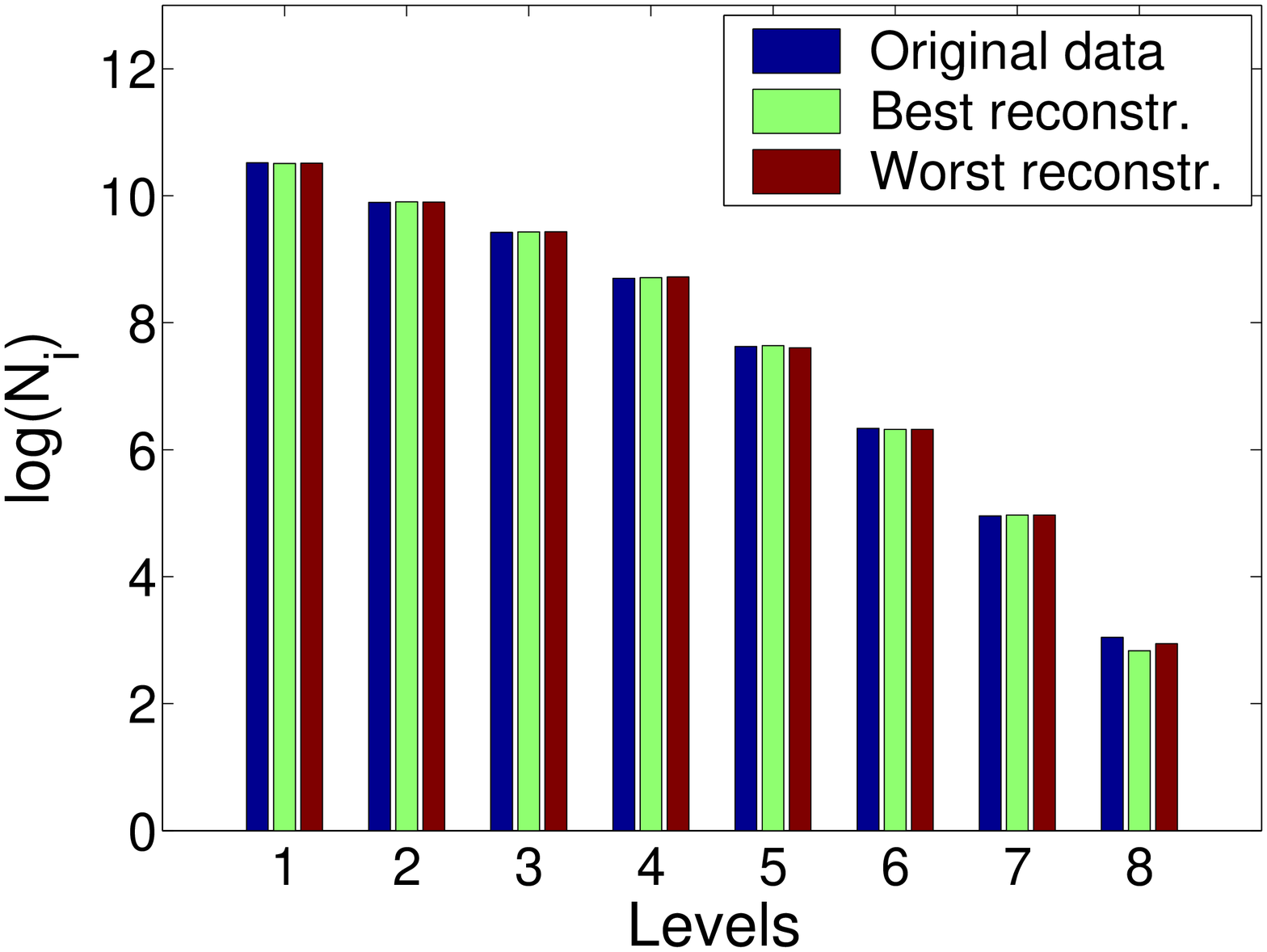}}
    \subfigure[P-NN-MLC, $p=0.66$]{\label{fig:hist-q16-msng66}
    \includegraphics[scale=0.33]{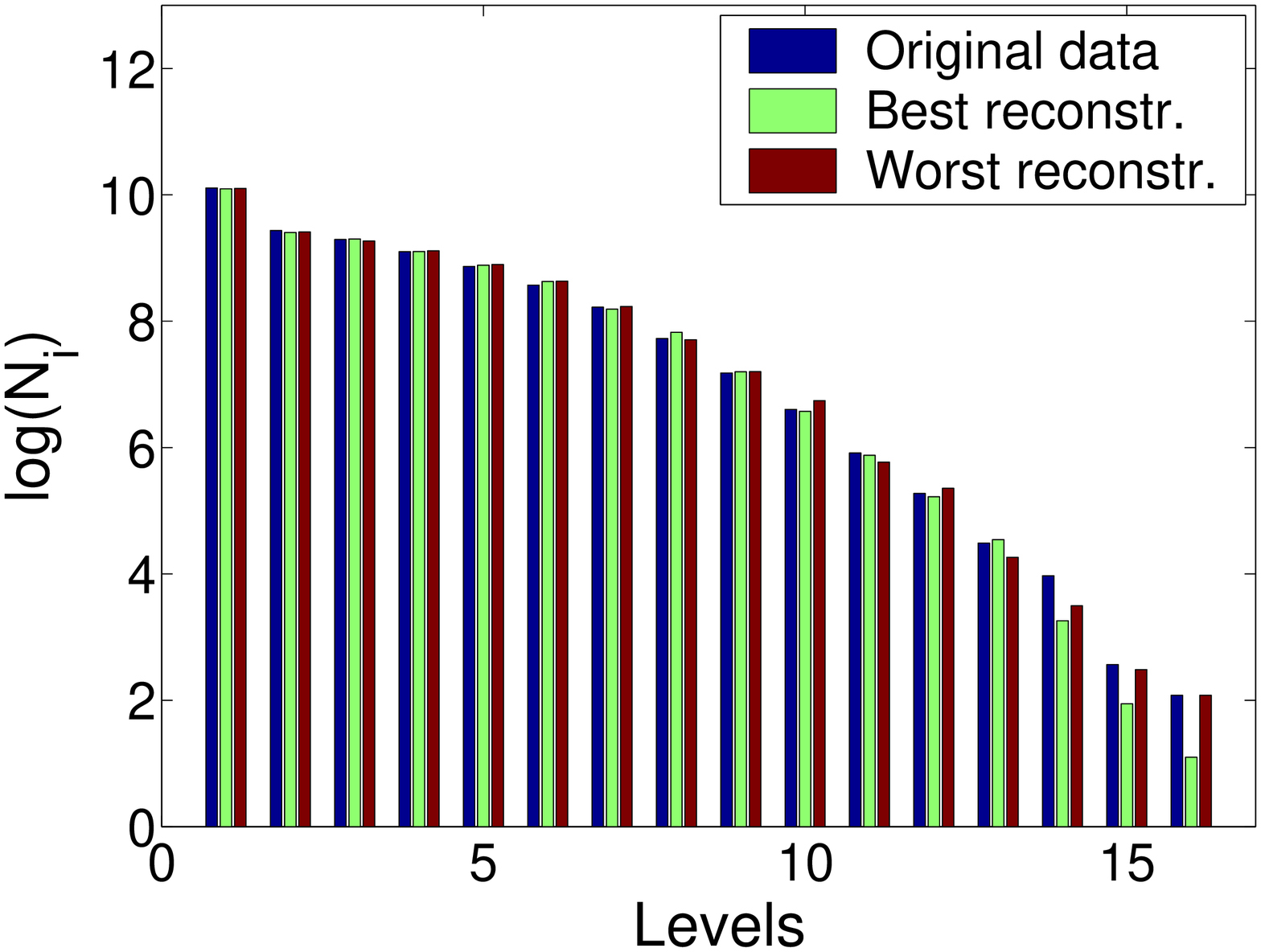}}
  \caption{The log-histograms of the original, the best $(F^{*}_{min}),$
  and the worst $(F^{*}_{max})$ reconstructed data, obtained by
  (a) I-NN-MLC with $p=0.33$, (b) I-NN-MLC with $p=0.66$,
  (c) P-NN-MLC with $p=0.33$, and (d) P-NN-MLC with $p=0.66$.}
  \label{fig:rec_hist}
\end{figure}

\begin{figure}[!ht]
    \subfigure[I-NN-MLC, $p=0.33$]{\label{fig:variox-M8-msng33}
    \includegraphics[scale=0.33]{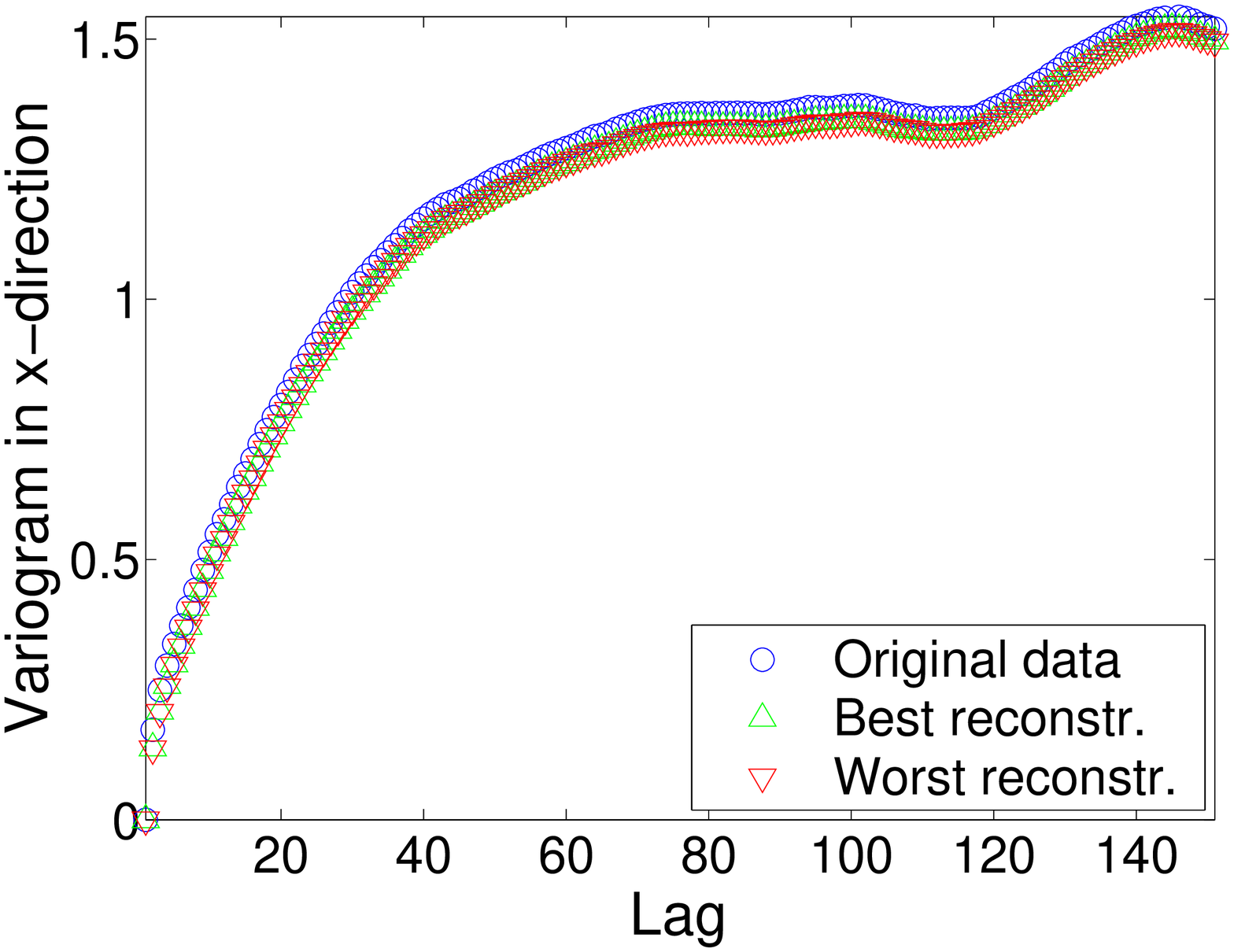}}
    \subfigure[I-NN-MLC, $p=0.66$]{\label{fig:variox-M16-msng66}
    \includegraphics[scale=0.33]{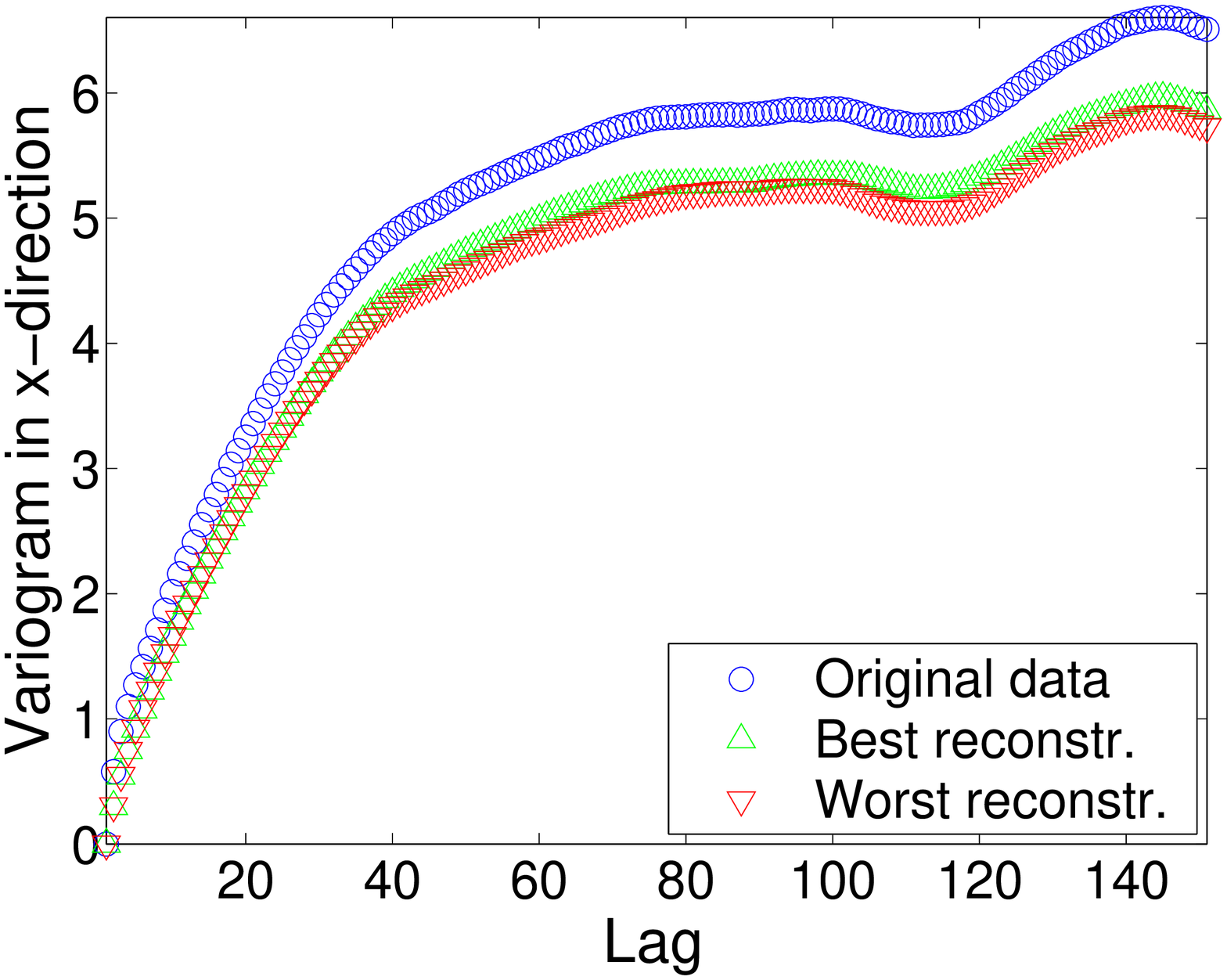}}\\
    \subfigure[P-NN-MLC, $p=0.33$]{\label{fig:variox-q8-msng33}
    \includegraphics[scale=0.33]{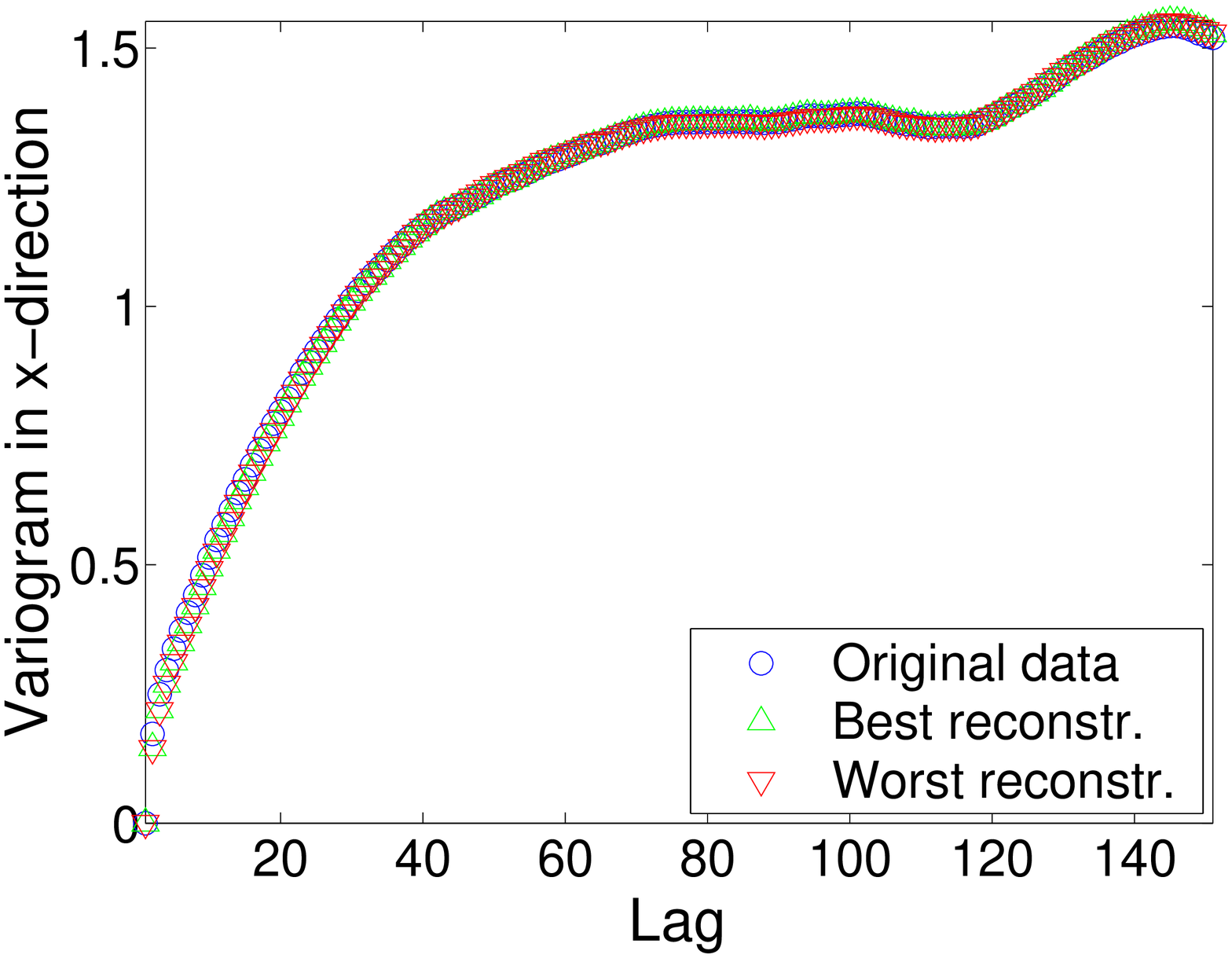}}
    \subfigure[P-NN-MLC, $p=0.66$]{\label{fig:variox-q16-msng66}
    \includegraphics[scale=0.33]{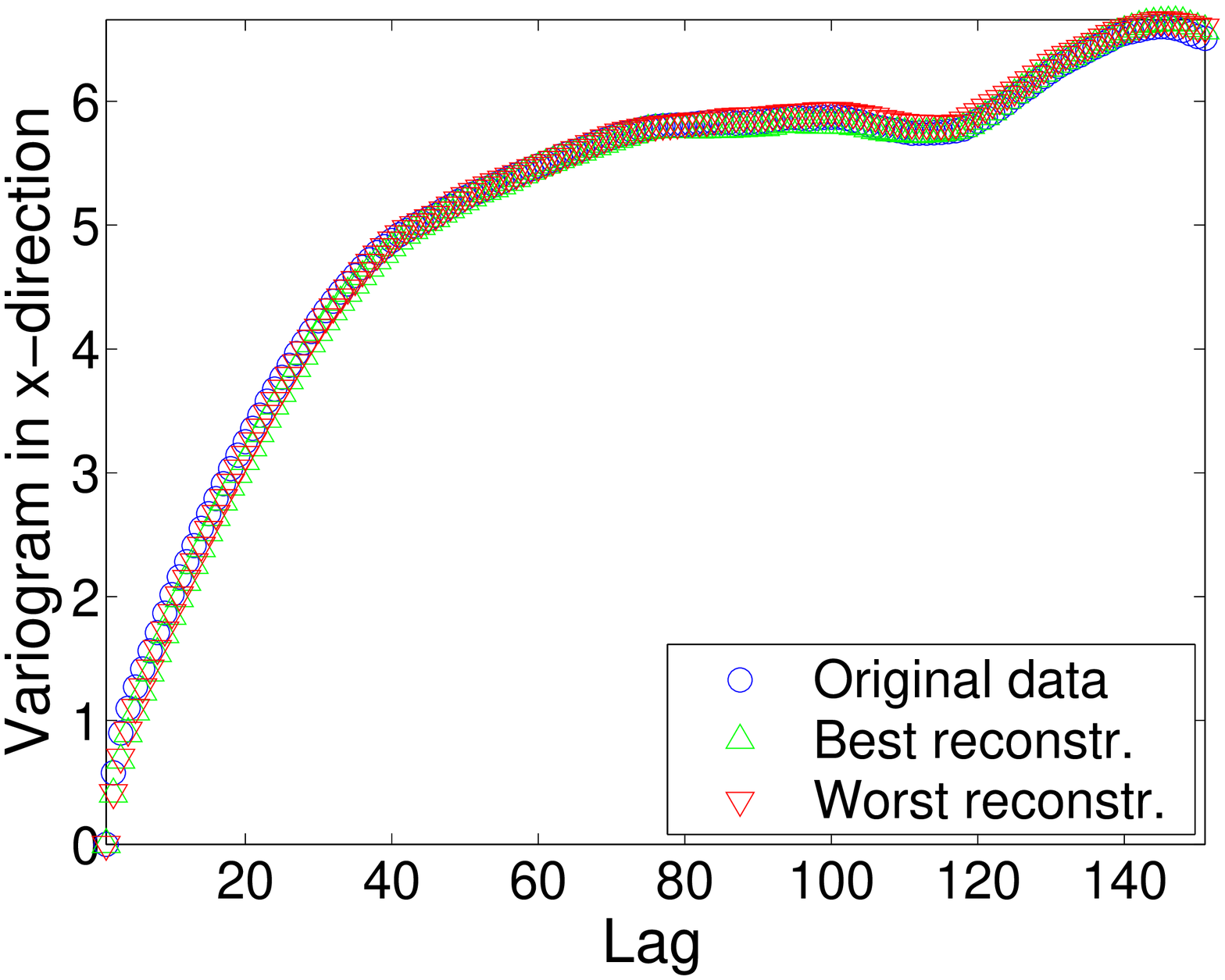}}
  \caption{The x-axis direction variograms of the original, the best
  $(F^{*}_{min}),$ and the worst $(F^{*}_{max})$ reconstructed data,
  obtained by (a) I-NN-MLC with $p=0.33$, (b) I-NN-MLC with $p=0.66$,
  (c) P-NN-MLC with $p=0.33$, and (d) P-NN-MLC with $p=0.66$.}
  \label{fig:rec_variox}
\end{figure}

\begin{figure}[!ht]
    \subfigure[I-NN-MLC, $p=0.33$]{\label{fig:varioy-M8-msng33}
    \includegraphics[scale=0.33]{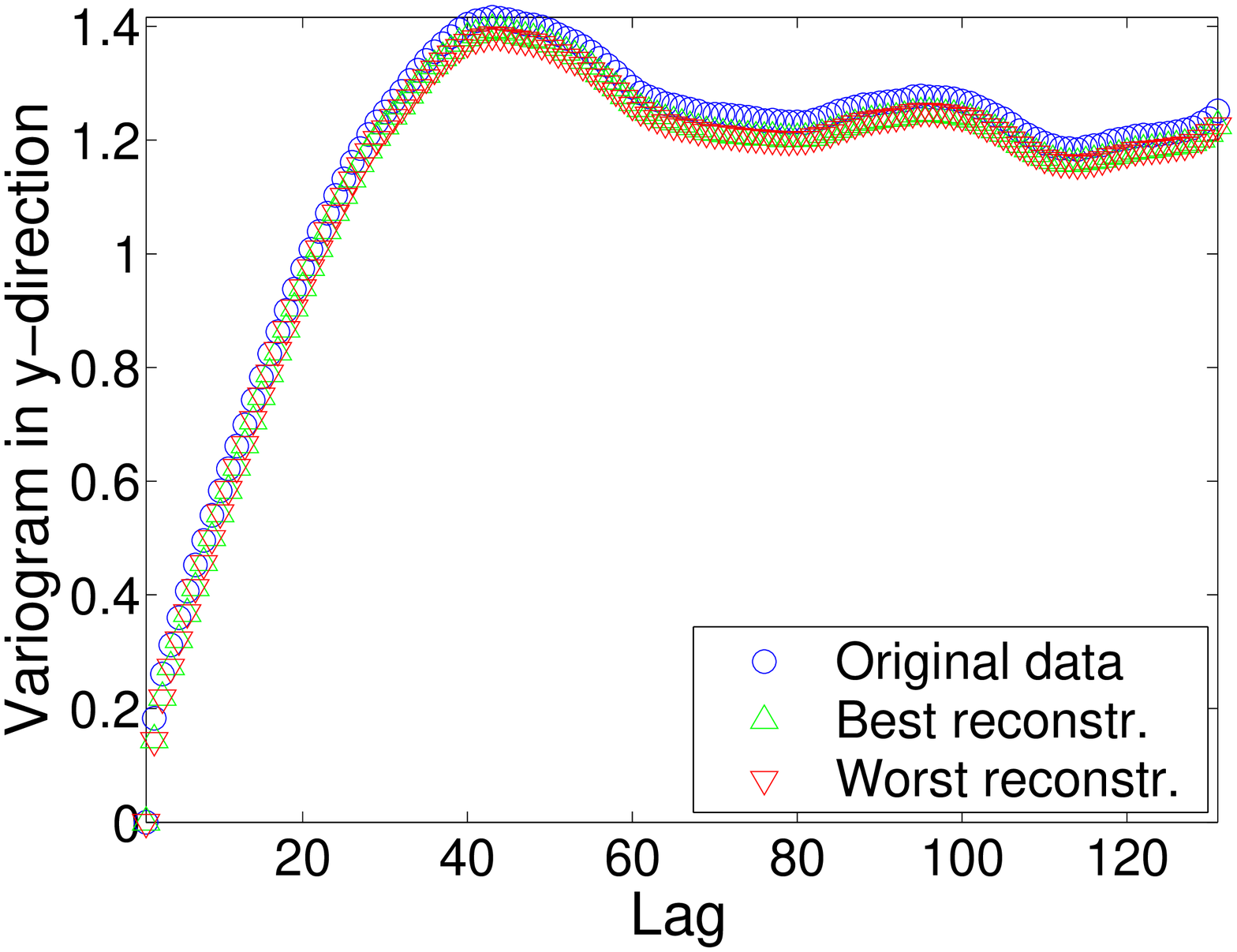}}
    \subfigure[I-NN-MLC, $p=0.66$]{\label{fig:varioy-M16-msng66}
    \includegraphics[scale=0.33]{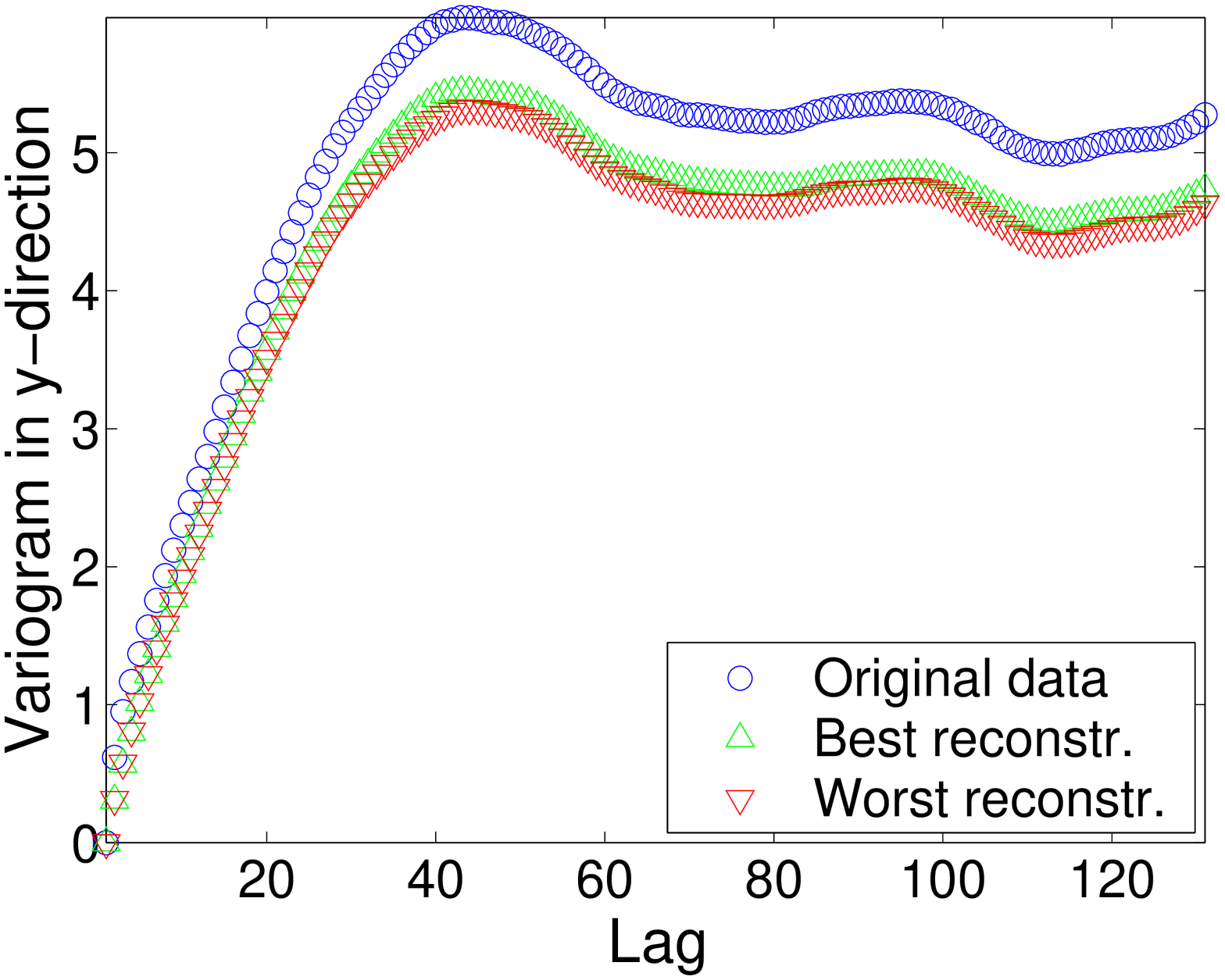}}\\
    \subfigure[P-NN-MLC, $p=0.33$]{\label{fig:varioy-q8-msng33}
    \includegraphics[scale=0.33]{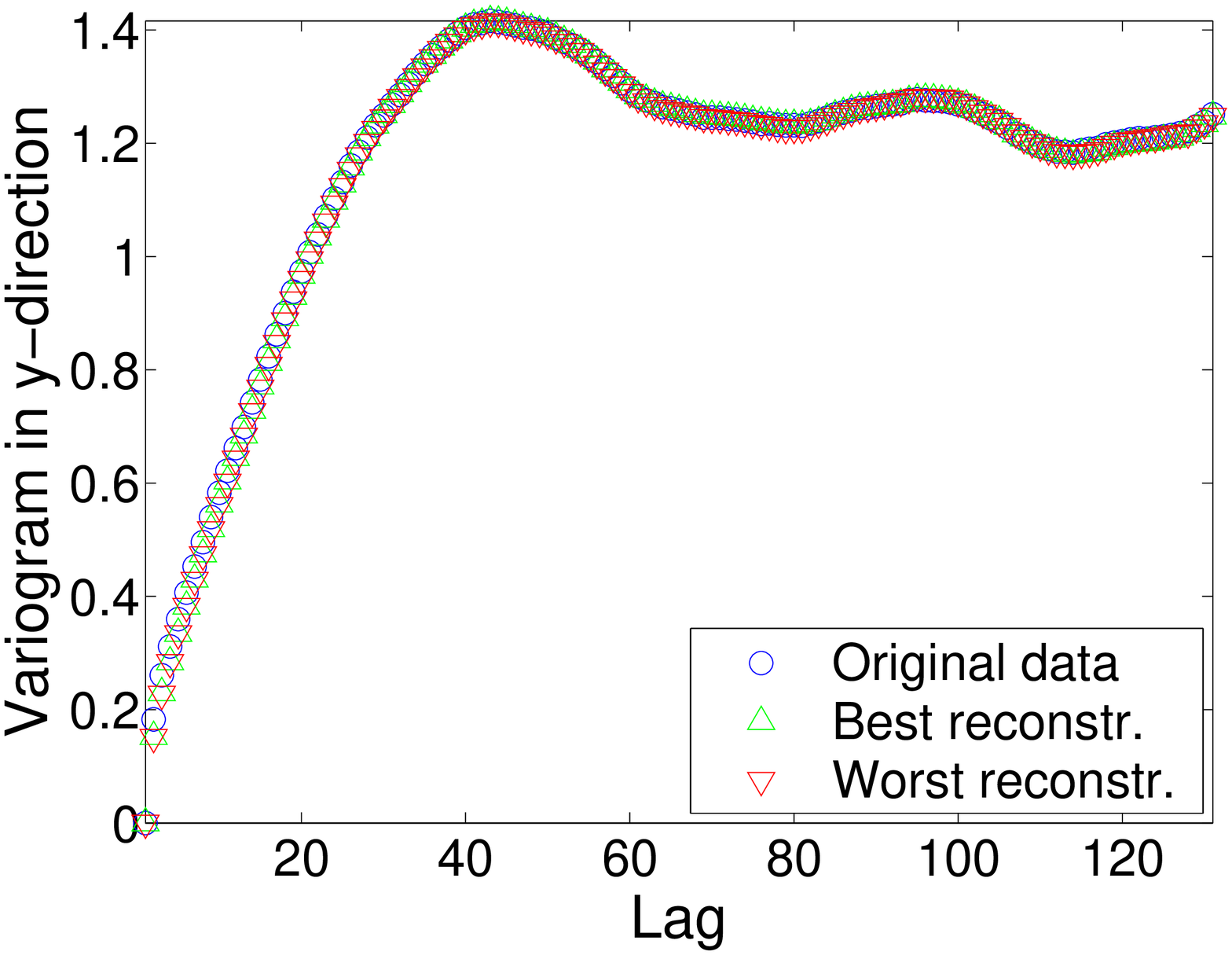}}
    \subfigure[P-NN-MLC, $p=0.66$]{\label{fig:varioy-q16-msng66}
    \includegraphics[scale=0.33]{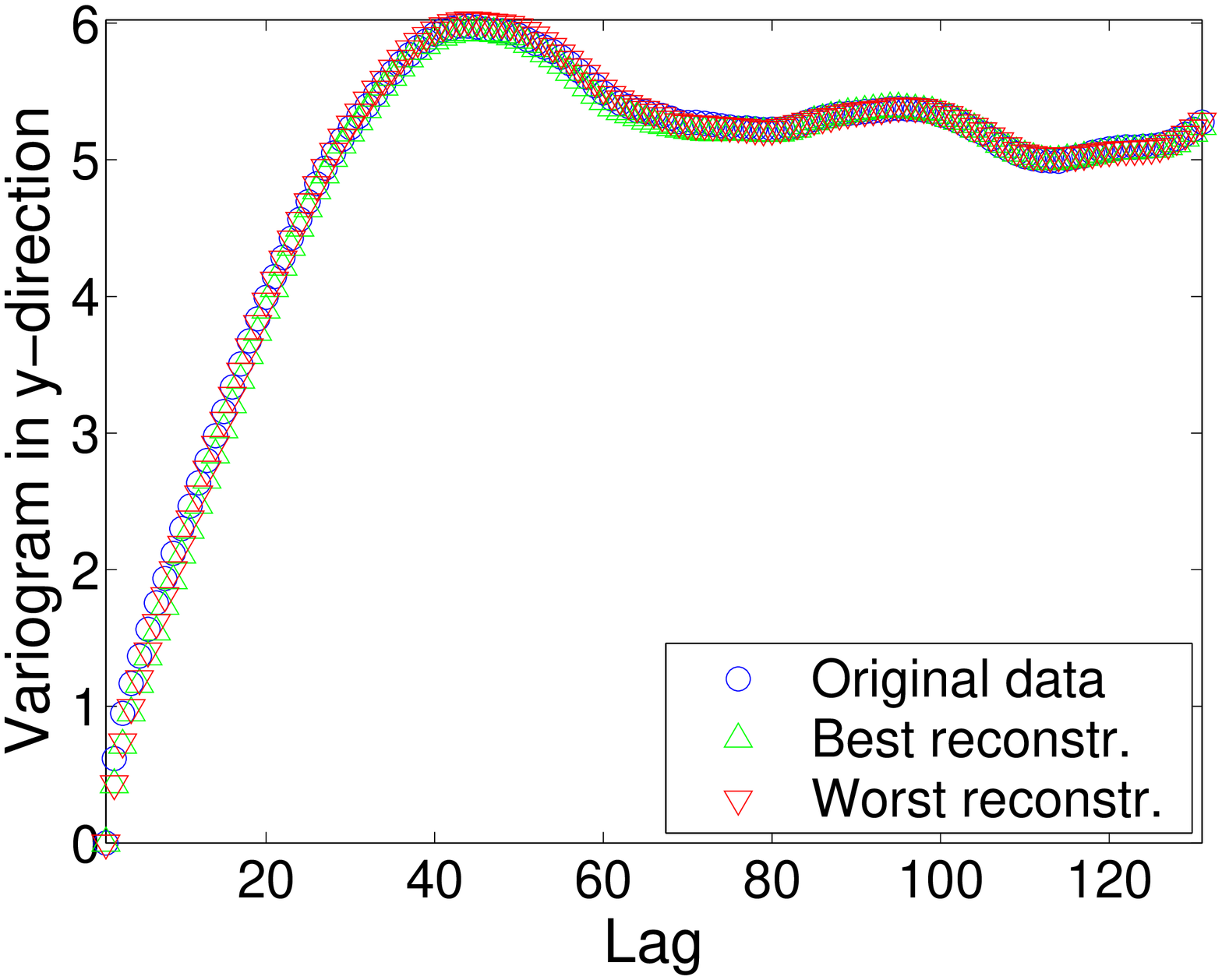}}
  \caption{The y-axis direction variograms of the original, the best
  $(F^{*}_{min}),$ and the worst $(F^{*}_{max})$ reconstructed data,
  obtained by (a) I-NN-MLC with $p=0.33$, (b) I-NN-MLC with $p=0.66$,
  (c) P-NN-MLC with $p=0.33$, and (d) P-NN-MLC with $p=0.66$.}
  \label{fig:rec_varioy}
\end{figure}

\subsection{Analysis of Missing Data Reconstruction Results}
 In Figs. \ref{fig:rec_maps} - \ref{fig:rec_varioy} the
classification performance of the two spin-based models is demonstrated
for two limiting cases: one with a low number of levels and low degree of thinning
($N_c=8$ and $p=0.33$), and the other one with a high number of levels and high
degree of thinning ($N_c=16$ and $p=0.66$). The plots in these figures
include: in Fig. \ref{fig:rec_maps}, a 2D projection of the reconstructed isolevel map (shown
for the first realization), in Fig. \ref{fig:rec_std}, the spatial distribution of the class
index standard deviations based on the $100$ realizations, in Fig. \ref{fig:rec_hist}, the
histograms of the original data as well as the best (lowest $F^*$)
and worst (highest $F^*$) reconstructions, and in Figs. \ref{fig:rec_variox} and
\ref{fig:rec_varioy}, the variograms (along the directions of coordinate axes) of  the original
data versus those of the best and worst reconstructions. In all
cases, there is good visual agreement between the spatial patterns
of the original data and the reconstructions. However, at higher
values of $p=0.66$ and $N_c=16$, closer inspection of the maps can
reveal some degree of smoothing and small speckles of misclassified pixels.
The higher misclassification in the $p=0.66$ and $N_c=16$ case
is also manifested in the remaining quantities, in particular, poorer matching of
the histograms, and larger deviations of the reconstructions' variogram curves
with respect to the original.

The histograms are displayed in log-lin scale in order to better
visualize also the small frequency classes (in the tail). The
natural logarithm of the class frequencies, $N_{i}$
$(i=1,\ldots,N_{c})$ is used. On the other hand, the logarithmic
scale somewhat visually suppresses the differences in the high
 frequency classes. Nevertheless, in the I-NN-MLC model we can observe
a systematic underestimation of the highest-frequency (first) class and the low-frequency classes
(especially at larger $p$). Their underestimation (and overestimation
of the classes closer to the mean $\bar{I}_Z^{16}\approx 3.3$) is
reflected in the noticeable decrease in the class index variance
of the reconstructed maps (as shown by the variogram plots). On the other hand, for the
P-NN-MLC model, the frequencies of the most represented classes are reconstructed reasonably
well, and the classes in the tail only represent a small portion of the total data
(e.g., for $N_c=16$ the classes larger than 13 represent less than 0.1\% of $N_{\tilde{G}}$) and,
therefore, the variation in their frequencies have relatively little impact on the variograms.

\begin{table}[h]
\begin{small}
\caption{The mean values of the misclassification rate  $\langle
F^{*} \rangle$ and the standard deviations ${\rm STD}_{F^{*}}$
obtained by the I-NN-MLC and P-NN-MLC models are compared with the
best results obtained by the $k$-NN classification ($\langle
F^{*}_{knn} \rangle$ and ${\rm STD}_{F^{*}_{knn}}$). The additional
statistics for the I-NN-MLC and P-NN-MLC models include: the mean
numbers of Monte Carlo sweeps $\langle N_{MC} \rangle$, the mean
values of the CPU time $\langle T_{cpu} \rangle$, and the mean
values of the cost function at termination $\langle U^{*} \rangle$.
The averaging is performed over 100 realizations.}
\label{tab:data-8}
\begin{center}
\begin{tabular}{lcccccc|cccccc}
\hline \hline
\# of classes    & \multicolumn{6}{c}{8 classes}  & \multicolumn{6}{c}{16 classes} \\
\hline
$p$        & $0.33$ & $0.5$ & $0.66$ & $0.33$ & $0.5$ & $0.66$ & $0.33$ & $0.5$ & $0.66$ & $0.33$ & $0.5$ & $0.66$\\
\hline
Model          & \multicolumn{12}{c}{k Nearest Neighbors}  \\
\hline
$\langle F^{*}_{knn} \rangle$ [\%]&     22.6 &    24.0  &    25.9  &     22.6 &     24.0 &    25.9  &   39.7  &     41.4 &     43.4 &     39.7 &     41.4 &     43.4\\
${\rm STD}_{F^{*}_{knn}}$ &   0.24 &    0.20 &     0.16 &    0.24 &    0.20 &   0.16   &    0.31 &    0.22 &     0.18 &    0.31 &   0.22 &    0.18\\
\hline
Model          & \multicolumn{3}{c}{I-NN-MLC}  & \multicolumn{3}{c}{P-NN-MLC} & \multicolumn{3}{c}{I-NN-MLC}  & \multicolumn{3}{c}{P-NN-MLC}\\
\hline
$\langle F^{*} \rangle$ [\%]&     21.4 &     22.6 &     24.2 &     21.7 &     22.9 &     24.5  &     37.1&     39.0 &     41.7 &     38.1 &     39.8 &     41.8\\
${\rm STD}_{F^{*}}$ &    0.23 &    0.19 &     0.17 &    0.23 &    0.20 &     0.17   &    0.28 &    0.22 &     0.21 &    0.29 &    0.21 &     0.17\\
$\langle N_{MC} \rangle$ &     5.9 &    7.5 &    9.1 &     31.8 &    35.9 &   41.4  &     13.8 &    17.4 &    19.4 &    65.8 &    74.6 &    85.0 \\
$\langle T_{cpu} \rangle$ [s]&    2.62 &     3.21   &    3.87 &     1.97 &     3.04   &   4.28   &      5.31 &     6.29   &    7.15 &     3.37 &    5.31   &    7.45 \\
$\langle U^{*} \rangle$ &    5e$-4$ &    1e$-3$ &   2e$-3$ &    1e$-3$ &    2e$-3$ &   2e$-3$  &    4e$-4$ &    8e$-4$ &   1e$-3$ &    3e$-3$ &     6e$-3$ &   6e$-3$ \\
\hline \hline
\end{tabular}
\end{center}
\end{small}
\end{table}

The misclassification rate and its standard deviation of the two
algorithms are compared in Table~\ref{tab:data-8}. In terms of the
misclassification rate, the I-NN-MLC model performs uniformly better
than the P-NN-MLC model. For the current set the differences are not
large, but they can be significant for different data (see
\ref{ssec:gaussian} below).  Comparing the proposed spin models with
the $k$-NN classifier, both models gave uniformly smaller
misclassification rates than the best achievable by the $k$-NN
algorithm.

\subsection{Computational performance of classification methods}
The computational performance of the proposed spin classifiers is
compared to the $k-$NN classifier also in Table~\ref{tab:data-8}.
 Due to binary values of the Ising spins, the I-NN-MLC model requires
a very small number of Monte Carlo sweeps over the grid to reach
equilibrium. In the most ``difficult'' case ($N_c=16$ and $p=0.66$),
it takes less than $20$ lattice sweeps. This implies short
optimization CPU times at each level. A substantial fraction of the
total CPU time is spent for the initial state assignments at each
level. For the P-NN-MLC model the initial state is determined once.
On the other hand, due to the significantly larger configuration space, the
relaxation is much slower than for the I-NN-MLC model. Nevertheless,
it is accomplished within maximum $32$ (fastest) and $85$ (slowest)
MC sweeps in less than $2$ (fastest) and $8$ (slowest) seconds of
CPU time. Optimizing the $k$-NN algorithm involved time-consuming
multiple runs for each realization to test a wide range of  $k$
values, leading to considerably higher CPU times (not reported). In
practical applications, the optimal value of $k$ is often selected
by heuristic techniques (e.g., cross-validation), which also require
user input and computational resources. Overall, the I-NN-MLC and
P-NN-MLC models can provide better classification accuracy more
efficiently and without user intervention.

\subsection{Reconstruction of synthetic Gaussian random field}
\label{ssec:gaussian} 
\begin{figure}[b]
  \begin{center}
    \includegraphics[width=0.48\textwidth]{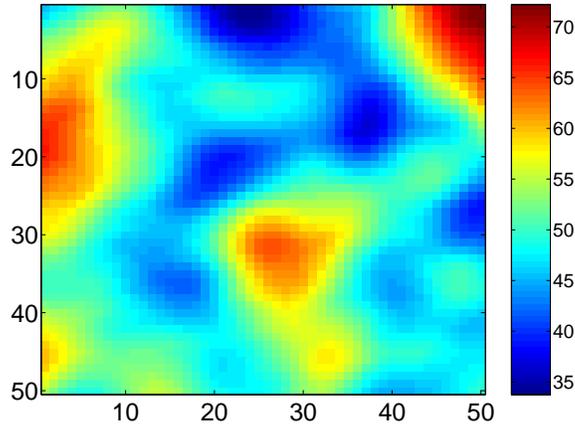}
  \end{center}
  \caption{Synthetic random field with a Gaussian distribution $Z \sim N(m=50,\sigma=10)$ and
  Whittle-Mat\'{e}rn type correlations ($\nu=2.5$ and $\kappa=0.2$).}
  \label{fig:mate}
\end{figure}

To further investigate differences in the
classification performance between the I-NN-MLC and P-NN-MLC models,
we generated smooth synthetic data on a $50\times50$ grid. The data
represent a realization  (see Fig. \ref{fig:mate}) from a Gaussian
random field $Z \sim N(m=50,\sigma=10)$ with Whittle-Mat\'{e}rn
correlations~\cite{whittle}.
The correlation function is $c (r)=\sigma^{2} \,
\frac{{2}^{1-\nu}}{\Gamma(\nu)}(\kappa r)^{\nu}K_{\nu}(\kappa r),$
where $\nu=2.5$ is the smoothness parameter, $\kappa=0.2$ is the
inverse correlation length, and $K_{\nu}$ is the modified Bessel
function of order $\nu$. The biggest difference in classification
accuracy between the two models is obtained for $N_c=16$ and
$p=0.33$: $F^{*}_{\rm{P-NN-MLC}}=35.1\%$, while
$F^{*}_{\rm{I-NN-MLC}}=21.2 \%$. We believe that the superior
performance of the I-NN-MLC model results from the sequential
strategy, in which points classified as $-1$ at lower levels are
included in the sampling set at higher levels. Provided that the
classification at lower levels is accurate, which is more likely for
rather smooth and noise-free data (like the synthetic ones shown
above), the included estimates can significantly improve the model's
performance. The sequential algorithm also reduces potential state
degeneracy (i.e., spin configurations with the same energy). This
feature is likely to occur in the spin models and it increases
ambiguity in the identification of a particular spin configuration
from the correlation energy. At the same time,  the propagation of
classification results from lower to higher levels can also be a
weakness of the sequential algorithm, since low level classification
errors  influence the higher levels.

\section{Conclusions and future research}
\label{conclusions}

We presented  non-parametric approaches for spatial classification,
inspired from the Ising and the Potts spin models, with a sequential
and simultaneous classification strategy, respectively. The concept
is based on the idea of matching the normalized correlation energies
calculated from discretized data over the sampling grid and the
entire area of interest. The matching is performed using Monte Carlo
simulations, conditional on the sample values. The main advantage of
the models is that they do not have any hypeparameters that need
tuning; hence the classification is automatic, objective,
competitive (in accuracy) and computationally efficient. The
proposed methods are applicable to non-Gaussian distributions. In
addition, they can incorporate non-stationary data, because even
with a constant coupling strength the spin interactions imply a
local impact of the sample values.

The future research includes the extension to scattered sampling
patterns. One possible way is to define the interaction constant
$J_{ij}$ in the Hamiltonians (\ref{Ising}-\ref{Potts}) through a
kernel function (such as the radial basis function), and the
interaction neighborhood (nearest neighbors), for example, as pairs
of points whose Voronoi cells have a common boundary. Another way is
to first use a simple interpolation method to place the irregularly
spaced points on a regular grid with a specified resolution and then
proceed as in the current study. The latter approach would allow
vectorization and preserve the computational efficiency. Further
possible extensions of the current models could also include
further-neighbor or/and ``higher-order'' (e.g., three-point)
correlation energy in the respective Hamiltonians. We could expect
some elimination of the degeneracy, witnessed in the present models,
and more faithful characterization of the nature of the spatial
dependance. Both effects should contribute to the improvement of the
classification performance. It would also be interesting to consider
data sets with different patterns of missing data and investigate
the effect of various gap patterns. Finally, it remains to be seen
if the proposed methods can be used in the case of data sets
containing a small number of extreme values, for example, two or
three unusually elevated values detected by a monitoring network in
the case of a radioactivity release.

\ack
 This research project has been supported by a Marie
Curie Transfer of Knowledge Fellowship of the European Community's
Sixth Framework Programme under contract number MTKD-CT-2004-014135.

\end{document}